\documentclass[preprint2]{aastex}
\usepackage{color}
\usepackage{epsfig}
\usepackage[fleqn]{amsmath}
\usepackage{multirow}
\usepackage{mathptmx}
\usepackage{graphicx}
\usepackage{bm}

\usepackage{lineno}
\shorttitle{Construction of explicit symplectic integrators}
\shortauthors{Wu et al.}

\begin{document}


\title{Construction of explicit symplectic integrators in general relativity. IV. Kerr black holes}


\author{Xin Wu$^{1,2,3,\dag}$, Ying Wang$^{1,2}$, Wei Sun$^{1,2}$, Fuyao Liu$^{1}$}
\affil{1. School of Mathematics, Physics and Statistics, Shanghai
University of Engineering Science, Shanghai 201620, China
\\ 2. Center of Application and Research of Computational Physics,
Shanghai University of Engineering Science, Shanghai 201620, China
\\  3. Guangxi Key Laboratory for Relativistic Astrophysics, Guangxi
University, Nanning 530004, China} \email{Emails:  $\dag$
Corresponding Author: wuxin$\_$1134@sina.com (X. W.),
wangying424524@163.com (Y. W.), sunweiay@163.com (W. S.),
liufuyao2017@163.com (F. L.)}
%\email{wangying424524@163.com} sunweiay@163.com, liufuyao2017@163.com


\begin{abstract}

In previous papers, explicit symplectic integrators were designed for nonrotating black holes, such as a Schwarzschild black hole. However, they fail to work in the Kerr spacetime because not all variables can be separable, or not all splitting parts have analytical solutions as explicit functions of proper time. To cope with this difficulty, we introduce a time transformation function to the Hamiltonian of Kerr geometry so as to obtain a time-transformed Hamiltonian consisting of five splitting parts, whose analytical solutions are explicit functions of the new coordinate time. The chosen time transformation function can cause time steps to be adaptive, but it is mainly used to implement the desired splitting of the time-transformed Hamiltonian. In this manner, new explicit symplectic algorithms are easily available. Unlike Runge–Kutta integrators, the newly proposed algorithms exhibit good long-term behavior in the conservation of Hamiltonian quantities when appropriate fixed coordinate time steps are considered. They are better than same-order implicit and explicit mixed symplectic algorithms and extended phase-space explicit symplectic-like methods in computational efficiency. The proposed idea on the construction of explicit symplectic integrators is suitable for not only the Kerr metric but also many other relativistic problems, such as a Kerr black hole immersed in a magnetic field, a Kerr-Newman black hole with an external magnetic field, axially symmetric core-shell systems, and five-dimensional black ring metrics.
\end{abstract}


\emph{Unified Astronomy Thesaurus concepts}: Black hole physics
(159); Computational methods (1965); Computational astronomy
(293); Celestial mechanics (211)

%\keywords{Computational methods --- Computational astronomy}



\section{Introduction}
\label{sec:intro}

Gravitational waves and black holes are two fundamental
predictions of the theory of general relativity of Einstein. The
predictions have been frequently confirmed by a number of
observations of gravitational waves from binary black hole or
neutron star mergers [e.g., GW150914 (Abbott et al. 2016) and
GW190521 (Abbott et al. 2020)] and images of the supermassive
black hole candidate in the center of the giant elliptical galaxy
M87 (EHT Collaboration et al. 2019). The observed image provides
powerful evidence for the presence of a rotating black hole, which
was derived by Kerr (1963) from the field equations of general
relativity. There are also other  solutions for the field
equations, such as a non-rotating Schwarzschild black hole.

The geodesics of the Schwarzschild, Reissner-Nordstr\"{o}m,
Reissner-Nordstr\"{o}m-(anti)-de Sitter and Kerr spacetime
geometries are highly nonlinear but are integrable. This
integrability does not mean that their solutions can be expressed
in terms of elementary functions but quadratures. Thus, numerical
integration schemes are still necessarily used to solve these
geodesic equations. The popular fourth-order Runge-Kutta explicit
integration method (RK4) is applicable for arbitrary metrics
(i.e., arbitrary spacetimes in arbitrary coordinates). Recently,
the RAPTOR code with RK4 was applied to study accretion models of
supermassive Kerr black holes so as to produce physically accurate
images of black-hole accretion disks (Bronzwaer et al. 2018,
2020). The RK4 integrator is an accurate and efficient integrator
for a short integration time. However, it would yield a secular
drift in energy errors and would provide unreliable numerical
results for a long-term integration. The manifold correction of
Nacozy (1971) and its extensions (Fukushima 2003; Wu et al. 2007;
Ma et al. 2008; Wang et al. 2016; Wang et al. 2018; Deng et al.
2020) are helpful to compensate for the defect of RK4 by pulling
the integrated orbit back to the original integral hypersurface.

The RK4 integrator combined with a manifold correction scheme is
regarded as one of the geometric integration algorithms that
preserve structures, integrals, symmetries, reversing symmetries,
and phase-space volumes (Hairer et al. 1999). The manifold
correction scheme can strictly satisfy energy integral, but does
not conserve the symplecticity. A class of discrete Hamiltonian
gradient schemes that conserve energy to machine precision can be
used for any globally hyperbolic spacetimes with six dimensions
(Bacchini et al. 2018, 2019), and eight- and ten-dimensional
Hamiltonian problems (Hu et al. 2019, 2021). These
energy-conserving discrete gradient integrators are implicit,
nonsymplectic, and do not preserve other integrals in
general\footnote{For example, the norm of a one-dimensional
disordered discrete nonlinear Schr\"{o}dinger equation is not
conserved by the energy-conserving integrator for ten-dimensional
conservative Hamiltonian systems (Hu et al. 2021). However, Figure
6 of Bacchini (2018) shows that the conservation of the Carter
constant for unstable spherical photon orbits around a Kerr Black
Hole is achieved to machine precision by the Hamiltonian discrete
gradient scheme.}. Their constructions become complex as the
dimensionality increases. No higher-order accuracy but only first-
and second-order accuracies can be given to numerical solutions in
the existing energy-conserving discrete gradient schemes.

Symplectic integrators (Swope et al. 1982; Wisdom 1982; Ruth 1983;
Forest $\&$ Ruth 1990; Wisdom $\&$ Holman 1991; Chambers $\&$
Murison 2000; Laskar $\&$ Robutel 2001; Omelyan et al. 2003) are
the best geometric integrators for studying the long-term
evolution of the geodesics and other Hamiltonian problems.
Although they are different from the manifold correction schemes
and energy-conserving integrators that can exactly preserve energy
integral, they remain bounded and show no secular growth in energy
errors. Other constants of motion and symplectic geometrical
structure are also conserved in the course of numerical
integrations. Because of the difficulty of the separation of
variables in curved spacetimes, the standard explicit symplectic
integrators become useless. Instead, implicit symplectic methods
(Feng 1986; Brown 2006; Tsang et al. 2015), or implicit and
explicit mixed symplectic methods (Preto $\&$ Saha 2009;
Kop\'{a}\v{c}ek et al. 2010; Lubich et al. 2010; Zhong et al.
2010; Mei et al. 2013a, 2013b) are used.

It is well known that an explicit algorithm is generally superior
to same order implicit method in computational efficiency. Noting
this fact, Pihajoki (2015) presented explicit leapfrog integration
schemes for inseparable Hamiltonian systems including arbitrary
metrics by doubling the phase space and introducing a new
Hamiltonian with two variable-separating parts equal to the
original Hamiltonian. Of course, the leapfrog algorithms are
symplectic in the extended phase space. However, the two parts are
coupled and dependent through the derivatives, and therefore, the
two numerical flows for the two parts diverge with time. Mixing
maps that act as feedback between the two solutions would address
the problem on the divergence of both solutions. If the mixing
maps are not symplectic, then the extended phase space leapfrogs
are not, either. Even if the mixing maps are symplectic, the
extended phase space leapfrogs are not symplectic when projection
maps are used to project a vector in extended phase space back to
the original phase space in any case. In spite of this, the
leapfrogs still preserve the original Hamiltonian without secular
growth in the error because the mixing maps and projection maps
are operated only in obtaining outputs and do not alter the state
in the extended phase space. In this sense,  these algorithms are
viewed as extended phase space explicit symplectic-like
integrators. Permutations of momenta were shown to be the best
mixing maps. Liu et al. (2016) proposed fourth-order extended
phase space explicit symplectic-like methods and found that the
sequent permutations of coordinates and momenta are superior to
the permutations of momenta. Luo et al. (2017) showed that the
best choice for permuted maps should be midpoint permutations.
These extended phase-space explicit symplectic-like integrators
are applicable for various inseparable problems (e.g., a vast
family of  spacetimes and post-Newtonian problems) (Li $\&$ Wu
2017; Luo $\&$ Wu 2017; Wu $\&$ Wu 2018; Li $\&$ Wu 2019). On the
other hand, Tao (2016) replaced the mixing maps with a third part
as an artificial restraint on the binding of the two copies of the
original system with mixed-up positions and momenta. In this way,
the extended phase-space leapfrogs are symplectic. More recently,
FANTASY based on this idea was applied to allow for the
integration of geodesics in arbitrary spacetimes with automatic
differentiation (Christian $\&$ Chan 2021). However, there is an
open problem on how to determine the most appropriate constant for
controlling the binding of the two copies. This choice is not
given in a universal method but relies on many numerical tests. Wu
$\&$ Wu (2018) reported that the Tao's method with an appropriate
choice of the control constant is not better than the method with
midpoint permutations in accuracy. Another problem is that the
extended phase-space leapfrogs are not symplectic when restricted
to the original phase space in any case, as Pihajoki claimed.

Are the standard explicit  symplectic  methods not applicable for
general relativistic metrics? No is the key to this question.
Recently, Wang et al. (2021a, Paper I) successfully separated
the Hamiltonian of Schwarzschild black hole into four integrable
parts with analytical solutions as explicit functions of proper
time, and used these explicit analytical solutions to compose the
standard second- and fourth-order explicit symplectic integrators.
When the Hamiltonian of a Reissner-Nordstr\"{o}m black hole has five
similar splitting parts, the standard explicit symplectic
integrators were easily set up in Paper II (Wang et al. 2021b).
The standard explicit symplectic integrators were also designed
for the Hamiltonian of a Reissner-Nordstr\"{o}m-(anti)-de Sitter
black hole with six integrable separable parts  in Paper III (Wang
et al. 2021c).

Unfortunately, the standard explicit symplectic integrators become
useless if the Hamiltonian of a Kerr black hole is split according
to the splitting techniques of the Hamiltonians of non-rotating
black holes in Papers I, II and III. This is because $\Sigma$ as a
function of $r$ and $\theta$ exists in the denominators of the
Hamiltonian, and leads to inseparable variables or splitting
Hamiltonian parts without the desired analytical solutions. To
overcome this difficulty, we use the time transformation method
introduced by Mikkola (1997) to obtain a time-transformed
Hamiltonian in which the denominators do not contain the function
$\Sigma$. In this way, the standard explicit symplectic algorithms
can be available for the time-transformed Hamiltonian. This is the
main aim of this paper.

The rest of this paper is organized as follows. In Sect. 2 we
introduce symplectic integrators with adaptive time steps of
Mikkola (1997). Then, the Kerr geometry is described in Sect. 3.
Explicit symplectic algorithms are designed for the Kerr geometry
in Sect. 4. We check the performance of the proposed algorithms in
Sect. 5.  Finally, the main results are concluded in Sect. 6. The
Carter constant and the parameters and initial conditions of
unstable spherical photon orbits in the Kerr spacetime are
described in Appendix A. Codes of the new second-order method are
given in Appendix B. Other choices of the time transformation function
are presented in Appendix C.

\section{Retrospect of symplectic integrators with adaptive time steps}

Consider a perturbed two-body problem with the Hamiltonian
\begin{equation}
H(\mathbf{p},\mathbf{r},\tau) = H_0 +R(\mathbf{r},\tau), ~~ H_0 =
\frac{\mathbf{p}^2}{2}-\frac{\mu}{r},
\end{equation}
where $\mathbf{p}$ denotes a momentum vector, $\tau$ is a physical
time, and $\mu$ is a constant associated with the constant of
gravity and masses of the two bodies.  The Kepler part $H_0$ is
integrable, and so is the perturbing function $R$.

Take $\tau=q_0$ as a new coordinate,  which corresponds to a
conjugate momentum $p_0=-H$. The phase space is extended by
$\mathbf{Q}=(q_0,\mathbf{r})$ and $\mathbf{P}=(p_0,\mathbf{p})$.
By introducing  a fictitious time variable $w$ through the
relation
\begin{equation}
d\tau=g(\mathbf{r},q_0)dw,
\end{equation}
Mikkola (1997) obtained an extended phase-space Hamiltonian
\begin{equation}
\Gamma=\Gamma_0+\Gamma_1, ~~ \Gamma_0=g(H_0+p_0), ~~ \Gamma_1=gR.
\end{equation}
$g$ and $\Gamma$ are called the time transformation function and
Hamiltonian, respectively.  $g$ is a constant along the orbit over
the fictitious time step from the beginning of each step to the
end of each step, but it changes at the beginning and at the end
of each step.  Even if $\Gamma$ is explicitly dependent on the
physical time coordinate $q_0$, it is identical to zero (i.e.
$\Gamma\equiv 0$) for any time $w$. $\Gamma_1$ does not depend on
any momenta and is thus easily solvable. However, $\Gamma_0$ is
difficult to analytically solve due to the inseparable
variables. To overcome this problem, Mikkola used the time
transformation $1/g$ (e.g. $g=r$) to transform $\Gamma_0$ back to
the physical time
\begin{equation}
\bar{H}_0 =\frac{1}{g}(\Gamma_0-\varepsilon)
=\frac{\mathbf{p}^2}{2}-\frac{\mu+\varepsilon}{r}+p_0,
\end{equation}
where $\varepsilon=\Gamma_0$ is the value of $\Gamma_0$ at the
beginning of the next step.  $\varepsilon$ is a constant along the
orbit over the physical time step during the beginning and the
end of each step. Equation (4) is still a Kepler problem with a
modified mass $\kappa=\mu+\varepsilon$. Of course, the mass at the
beginning of one step is unlike that at the end of this step.
Because $dw=d\tau/g=d\tau/r$, $h=w=\int^{\tau}(1/r)d\tau$ is a
time step. Based on the Stumpff's form of Kepler's equation, the
physical time step $\tau$ can be expressed in terms of $h$. The
positions and velocities of $\bar{H}_0$ are also functions of $h$.
Take $\mathcal{A}$ as a differential operator with respect to
$\bar{H}_0$. The analytical solutions (i.e. the momentum jumps) of
$\Gamma_1$ are easily obtained by $\Delta
\mathbf{p}=-h\partial(gR)/\partial \mathbf{r}$ and $\Delta
p_0=-h\partial(gR)/\partial q_0$. $\mathcal{B}$ is viewed as
another differential operator with respect to $\Gamma_1$. In this
way, Mikkola established the second-order symplectic leapfrog of
Wisdom $\&$ Holman (1991)
\begin{eqnarray}
    \mathcal{S}^{\Gamma}_2
    =\mathcal{B}(\frac{h}{2})\circ\mathcal{A}(h)\circ\mathcal{B}(\frac{h}{2}).
\end{eqnarray}
The operators $\mathcal{A}$ and $\mathcal{B}$ can also compose the
fourth-order symplectic method of Forest $\&$ Ruth (1990)
\begin{eqnarray}
    \mathcal{S}^{\Gamma}_4
    &=&\mathcal{B}(\frac{h}{2b_1})\circ\mathcal{A}(\frac{h}{b_1})\circ\mathcal{B}(\frac{1-b_2}{2b_1}h)\circ
    \mathcal{A}(-\frac{b_2}{b_1}h) \nonumber \\ &&
     \circ\mathcal{B}(\frac{1-b_2}{2b_1}h)\circ\mathcal{A}(\frac{h}{b_1})\circ\mathcal{B}(\frac{h}{2b_1}),
\end{eqnarray}
where $b_2=2^{1/3}$ and $b_1=2-b_2$. These adaptive integrators,
Equations (5) and (6), nearly preserve the Hamiltonian $\Gamma$, i.e., the
Hamiltonian (1). They demonstrate good qualitative properties of
symplectic integrators with constant time-steps. In addition, the
use of variable steps causes accuracy and efficiency to be
significantly improved.

The above symplectic algorithms are implemented in the new time
$w$. They rely on the analytical solutions of the Keplerian motion
$\bar{H}_0$ and require that the physical time $\tau$ should be
given in an expressional form of $h$. It is a hard task. However,
when the Hamiltonian (1) in the extended phase space is split into 
kinetic energy $T=\mathbf{p}^2/2+p_0$ and the potential energy
$U=R(\mathbf{r},q_0)$, such similar explicit symplectic
integrators are easily available for the logarithmic Hamiltonian
method proposed by Mikkola $\&$ Tanikawa (1999) and the time
transformation function suggested by Preto $\&$ Tremaine (1999).
Extensions and applications of such adaptive time step symplectic
integrators were considered by Mikkola $\&$ Aarseth (2002),
Emel'yanenko (2007), Preto $\&$ Saha (2009), Mikkola $\&$ Tanikawa
(2013), Ni $\&$ Wu (2014), Li $\&$ Wu (2017), and Wang $\&$
Nitadori (2020).

Since the fixed time step $h$ is used in the new fictitious time
$w$, the good long-time behavior properties of symplectic methods
are not lost. In addition, the physical time steps vary in
different positions of an orbit. In particular, they become
smaller as the perturbing body is closer to the central body. This
leads to increasing precision. The two points are what the
symplectic methods with time transformations satisfy.

\section{Kerr black hole}

The Kerr black hole is a rotating black hole. Its gravitational
field  is described by the spacetime metric
\begin{equation}
-d\tau^2 = ds^{2} =g_{\alpha\beta}dx^{\alpha}dx^{\beta}.
\end{equation}
In the standard Boyer-Lindquist coordinates $(t, r, \theta,
\phi)$, this metric is an axially symmetric stationary and has
 covariant nonzero components (Kerr 1963; Takahashi $\&$ Koyama 2009):
\begin{eqnarray}
&& g_{tt}=-(1-\frac{2r}{\Sigma}), ~~
g_{t\phi}=-\frac{2ar\sin^{2}\theta}{\Sigma}=g_{\phi t}, \nonumber \\
&&  g_{rr}=\frac{\Sigma}{\Delta}, ~~~~~~~~~~~~
g_{\theta\theta}=\Sigma,  \nonumber \\
&& g_{\phi\phi}=(\rho^2+\frac{2ra^2}{\Sigma}
\sin^{2}\theta)\sin^{2}\theta. \nonumber
\end{eqnarray}
Its contravariant nonzero components are
\begin{eqnarray}
&& g^{tt}=-\frac{A}{\Delta\Sigma}, ~~
g^{t\phi}=-\frac{2ar}{\Delta\Sigma}=g^{\phi t}, \nonumber \\
&&  g^{rr}=\frac{\Delta}{\Sigma}, ~~~~~~
g^{\theta\theta}=\frac{1}{\Sigma},  \nonumber \\
&& g^{\phi\phi}=\frac{\Sigma-2r}{\Delta\Sigma\sin^{2}\theta}.
\nonumber
\end{eqnarray}
Note that $\Sigma=r^2+a^2\cos^{2}\theta$, $\Delta=\rho^2-2r$,
$\rho^2=r^2+a^2$ and $A=\rho^4-\Delta a^2\sin^{2}\theta$. $\tau$
is a proper time, and $t$ is a coordinate time. The speed of light
$c$ and the gravitational constant $G$ use geometrized units,
$c=G=1$. The mass of the black hole $M$ also takes one unit,
$M=1$. In such a unit system, $a$ denotes the angular momentum of
the rotating body. The body should be slowly rotating, namely,
$|a|\leq 1$.

Based on the spacetime metric, a Lagrangian system is given by
\begin{equation}
\ell = \frac{1}{2} (\frac{ds}{d\tau})^2
=\frac{1}{2}g_{\mu\nu}\dot{x}^{\mu}\dot{x}^{\nu}.
\end{equation}
Four-velocity $\dot{x}^{\mu}=(\dot{t},\dot{r},
\dot{\theta},\dot{\phi})=U^{\mu}=\mathbf{U}$ or
$\mathbf{U}=U_{\mu}=g_{\mu\nu}U^{\nu}= g_{\mu\nu}\dot{x}^{\nu}$
satisfies the identical relation
\begin{equation}
\mathbf{U}\cdot \mathbf{U}=U_{\mu}U^{\mu}
=g_{\mu\nu}\dot{x}^{\mu}\dot{x}^{\nu}=-1.
\end{equation}
According to classical mechanics, a covariant generalized
four-momentum is defined as
\begin{equation}
p_{\mu} = \frac{\partial \ell}{\partial
\dot{x}^{\mu}}=g_{\mu\nu}\dot{x}^{\nu}.
\end{equation}
Clearly, each component of the metric tensor $g_{\mu\nu}$ does not
explicitly depend on the coordinates $t$ and $\phi$. Based on the
Euler-Lagrangian equations, the four-momentum exists two constant
components
\begin{eqnarray}
p_{t} = g_{tt}\dot{t}+g_{t\phi}\dot{\phi} &=&
-(1-\frac{2r}{\Sigma})\dot{t}
-\frac{2ra\sin^{2}\theta}{\Sigma}\dot{\phi} \nonumber \\
 &=& -E,\\
p_{\phi} = g_{\phi\phi}\dot{\phi}+g_{t\phi}\dot{t} &=&
(\rho^2+\frac{2ra^2}{\Sigma}
\sin^{2}\theta)\sin^{2}\theta\dot{\phi} \nonumber \\
&& -\frac{2ra\sin^{2}\theta}{\Sigma}\dot{t} =L.
\end{eqnarray}
$E$ represents the energy of a test particle moving around the
rotating body, and $L$ is the angular momentum of the particle.
Equations (11) and (12) can be rewritten as
\begin{eqnarray}
\dot{t} &=&\frac{dt}{d\tau}= -f_1E-f_2L, \\
\dot{\phi} &=& \frac{d\phi}{d\tau}= f_2E+f_3L,
\end{eqnarray}
where $f_1$, $f_2$ and $f_3$ are functions of $r$ and $\theta$ as
follows:
\begin{eqnarray}
f_1 &=& \frac{g_{\phi\phi}}{g_{tt}g_{\phi\phi}-g^{2}_{t\phi}}, \\
f_2 &=& \frac{g_{t\phi}}{g_{tt}g_{\phi\phi}-g^{2}_{t\phi}}, \\
f_3 &=& \frac{g_{tt}}{g_{tt}g_{\phi\phi}-g^{2}_{t\phi}}.
\end{eqnarray}

As in classical mechanics, this Lagrangian is strictly equivalent
to the Hamiltonian
\begin{eqnarray}
H &=& \mathbf{U}\cdot\mathbf{p}-\ell
=\frac{1}{2}g^{\mu\nu}p_{\mu}p_{\nu} \nonumber \\
&=& F +\frac{1}{2}\frac{\Delta}{\Sigma}p^{2}_{r}
+\frac{1}{2}\frac{p^{2}_{\theta}}{\Sigma},
\end{eqnarray}
where $F$ is a function of $r$ and $\theta$ and reads
\begin{eqnarray}
F &=&\frac{1}{2}(g^{tt}E^2 +g^{\phi\phi}L^2) -g^{t\phi}EL \nonumber \\
&=& \frac{1}{2}(f_1E^2 +f_3L^2)+f_2EL \nonumber \\
&=& -\frac{AE^2}{2\Delta\Sigma}+\frac{L^2(\Sigma-2r)}{2\Delta
\Sigma\sin^{2}\theta}+\frac{2ar}{\Delta\Sigma}EL.
\end{eqnarray}
The motion of the particle around the Kerr black hole is
determined by the Hamiltonian (18) with two degrees of freedom in
a 4-dimensional phase space. Besides the two constants $E$ and
$L$, the Hamiltonian itself is always equal to a known constant:
\begin{equation}
H=-\frac{1}{2}.
\end{equation}
This constant is due to the four-velocity relation (9).

It is clear that the Hamiltonian system (18) has two degrees of
freedom with four phase-space variables $(p_r,p_{\theta}; r,
\theta)$. Besides the Hamiltonian itself as an integral, a second
integral\footnote{Here, the constants $E$ and $L$ are excluded in
the integrals considered. They are only viewed as two dynamical
parameters of the system.} can be found by use of the separation
of variables in the Hamilton-Jacobi equations. This is the
so-called Carter constant (Carter 1968), which is given in
Appendix A. Thus, the system $H$ is integrable and formally
analytically solvable. In spite of this, no elementary functions
but quadratures can be given to the formal solutions. In this
case, it is still necessary to employ good numerical methods to
study the long-term evolution of geodesics of the Kerr geometry.

\section{Construction of explicit symplectic integrators}

In the previous papers (Wang et al. 2021a, 2021b, 2021c), we
designed explicit symplectic integrators for the Hamiltonians of
non-rotating black holes like the Reissner-Nordstr\"{o}m black
hole. This successful construction is completely based on the
splitting parts of each Hamiltonian that exists analytical
solutions as explicit functions of proper time $\tau$. If a
splitting technique similar to that of the Hamiltonian of
Reissner-Nordstr\"{o}m black hole in previous Paper II (Wang et
al. 2021b) is given to the Hamiltonian (18), we have
\begin{eqnarray}
H &=& H_1+H_2+H_3+H_4+H_5, \nonumber \\
H_1 &=& F, \nonumber \\
H_2 &=& \frac{1}{2}\frac{r^2}{\Sigma}p^{2}_{r}, \nonumber \\
H_3 &=& \frac{1}{2}\frac{a^2}{\Sigma}p^{2}_{r}, \nonumber \\
H_4 &=& -\frac{r}{\Sigma}p^{2}_{r}, \nonumber \\
H_5 &=& \frac{1}{2}\frac{p^{2}_{\theta}}{\Sigma}.
\end{eqnarray}
$H_1$ is easily solved. However, it is difficult to give
analytical solutions to any one of $H_2$, $H_3$, $H_4$ and $H_5$
because none of the Hamiltonians are separable to the variables.
Even if these sub-Hamiltonians can be solved analytically, but
their solutions are not explicit functions of proper time $\tau$.
For example, the evolution of $r$  with  $\tau$ in the
sub-Hamiltonian $H_3$ is described by
\begin{eqnarray}
\frac{r}{2}\sqrt{\Sigma}+\frac{a^2\cos^{2}\theta}{2}\ln(r+\sqrt{\Sigma})=c_1a^2\tau+c_2,
\end{eqnarray}
where $c_1$ and $c_2$ are integral constants. Because $r$ is only
one implicit function of $\tau$, the splitting Hamiltonian method
fails to construct an explicit symplectic integrator. In fact,
this failure is directly due to $\Sigma$ in the denominators of
the $p_r$ and $p_{\theta}$ terms acting as a function of $r$ and
$\theta$. To successfully construct explicit symplectic
integrators for the above-mentioned Hamiltonian, we must eliminate
the function $\Sigma$ in the denominators by constructing a
time-transformed Hamiltonian like Equation (3).

In the present case, we take the time transformation (2) in the
form
\begin{equation}
d\tau=g(r,\theta)dw,
\end{equation}
where $\tau$ is still the proper time and $w$ is a new coordinate
time unlike the original coordinate time $t$. Now the proper time
$\tau$ is referred to as a coordinate $q_0=\tau$, and its
corresponding momentum is $p_0$.  Note that $p_0\neq p_t$ but
$p_0=-H=1/2$. Then, the original phase-space variables
$(p_r,p_{\theta}; r, \theta)$ are extended to a set of new
phase-space variables $(p_0,p_r,p_{\theta}; q_0, r, \theta)$.
Similar to Equation (3), a time-transformed Hamiltonian is given
to the Hamiltonian (18) in the form
\begin{equation}
\mathcal{H}=g(H+p_0).
\end{equation}

When the time transformation function in Equation (23) takes
\begin{equation}
g(r,\theta)=\frac{\Sigma}{r^2},
\end{equation}
the time-transformed Hamiltonian in Equation (24) is
\begin{equation}
\mathcal{H}=\frac{\Sigma}{r^2} (F+p_0)
+\frac{\Delta}{2r^2}p^{2}_{r} +\frac{1}{2r^2}p^{2}_{\theta}.
\end{equation}
The denominators in the second and third terms of the new
Hamiltonian $\mathcal{H}$ just eliminate the function $\Sigma$,
compared with those in the original Hamiltonian $H$ of Equation
(18).

An operator-splitting technique is easily given to the Hamiltonian
$\mathcal{H}$. However, it is unlike that to the Hamiltonian
$\Gamma$ with two separable integrable parts in Equation (3). The
separable form of $\mathcal{H}$ is almost the same as that of the
Hamiltonian of Reissner-Nordstr\"{o}m black hole in previous Paper
II. The Hamiltonian (26) takes five splitting parts
\begin{equation}
\mathcal{H}=\mathcal{H}_1+\mathcal{H}_2+\mathcal{H}_3+\mathcal{H}_4+\mathcal{H}_5,
\end{equation}
where the sub-Hamiltonians $\mathcal{H}_i~(i=1,\ldots,5)$ are
\begin{eqnarray}
\mathcal{H}_1 &=& \frac{\Sigma}{r^2} (F+p_0), \\
\mathcal{H}_{2} &=& \frac{1}{2}p^{2}_{r},\\
\mathcal{H}_{3} &=& -\frac{1}{r}p^{2}_{r},\\
\mathcal{H}_{4} &=&
\frac{a^2}{2r^2}p^{2}_{r}, \\
\mathcal{H}_{5} &=& \frac{1}{2r^2}p^{2}_{\theta}.
\end{eqnarray}
The equations of motion for the sub-Hamiltonian $\mathcal{H}_1$ in
the new coordinate time read $dp_{0}/dw =dr/dw=d\theta/dw=0$, and
\begin{eqnarray}
\frac{d\tau}{dw} &=& \frac{\Sigma}{r^2} =g(r,\theta),   \\
\frac{dp_r}{dw} &=& \frac{2}{r}(\frac{\Sigma}{r^2}-1)(F+p_0)-
\frac{\Sigma}{r^2} \frac{\partial F}{\partial r}=P_{r}(r,\theta), \nonumber \\
\frac{dp_{\theta}}{dw} &=&
\frac{a^2}{r^2}(F+p_0)\sin(2\theta)-\frac{\Sigma}{r^2}
\frac{\partial F}{\partial \theta}=P_{\theta}(r,\theta). \nonumber
\end{eqnarray}
The equations of motion for the other sub-Hamiltonians are written
as
\begin{eqnarray}
\mathcal{H}_2: \frac{dr}{dw} &=& p_{r}, ~~\dot{p}_{r}=0; \\
\mathcal{H}_3: \frac{dr}{dw} &=& -\frac{2}{r}p_r, ~
\frac{d p_r}{dw} = -\frac{p^{2}_r}{r^2}; \\
\mathcal{H}_4: \frac{dr}{dw} &=& \frac{a^2}{r^2}p_r, ~
\frac{d p_r}{dw} = \frac{a^{2}}{r^3}p^{2}_r; \\
\mathcal{H}_5:  \frac{d\theta}{dw} &=& \frac{p_{\theta}}{r^{2}}, ~
\frac{dp_r}{dw} = \frac{p^2_{\theta}}{r^{3}}, ~
\dot{r}=\dot{p}_{\theta}=0.
\end{eqnarray}
Each equation is independently solved in an analytical method. Let
$(r_0,\theta_0,p_{r0},p_{\theta0})$ be the values at the beginning
of one step, and $(r,\theta,p_{r},p_{\theta})$ denote the
analytical solutions at the end of the step over time $w$. The
analytical solutions  are easily given to the five pieces, unlike
those in Equation (4). They are labeled as operators
$\tilde{\mathcal{H}}_{1}(w)$, $\tilde{\mathcal{H}}_{2}(w)$,
$\tilde{\mathcal{H}}_{3}(w)$, $\tilde{\mathcal{H}}_{4}(w)$ and
$\tilde{\mathcal{H}}_{5}(w)$. In fact, they are explicit functions
of the new coordinate time $w$:
\begin{eqnarray}
\tau(w) &=& \tau_0+ w g(r_0,\theta_0), \nonumber \\
\tilde{\mathcal{H}}_{1}: ~p_r(w) &=& p_{r0}+w P_{r}(r_0,\theta_0), \\
p_{\theta}(w) &=& p_{\theta0}+w P_{\theta}(r_0,\theta_0); \nonumber \\
\tilde{\mathcal{H}}_{2}: ~r(w) &=& r_0+w p_{r0}; \\
\tilde{\mathcal{H}}_{3}: ~ r(w) &=& [(r^{2}_{0}-3w
p_{r0})^{2}/r_0]^{\frac{1}{3}}, \nonumber \\
 p_r(w) &=& p_{r0}[(r^{2}_{0}-3w
p_{r0})/r^2_0]^{\frac{1}{3}}; \\
\tilde{\mathcal{H}}_{4}: ~ r(w) &=& (2a^{2}p_{r0}w/r_0+r^2_0)^{\frac{1}{2}}, \nonumber \\
 p_r(w) &=& \frac{p_{r0}}{r_0}(2a^{2}p_{r0}w/r_0+r^2_0)^{\frac{1}{2}}; \\
\tilde{\mathcal{H}}_{5}: ~ \theta(w) &=& \theta_0+w p_{\theta0}/r^{2}_{0}, \nonumber \\
 p_r(w) &=& p_{r0}+w p^2_{\theta0}/r^{3}_{0}.
\end{eqnarray}
Any one of the three compositions, involving Equations (29) and
(30), Equations (29)-(31), and Equations (29)-(32), is
analytically solvable. However, the analytical solutions are
implicit functions of $w$.

Using the splitting operators with a new coordinate time step $h$,
we design an explicit second-order symplectic integrator for the
system $\mathcal{H}$
\begin{eqnarray}
S^{\mathcal{H}}_2(h) &=& \tilde{\mathcal{H}}_5(\frac{h}{2})\circ
\tilde{\mathcal{H}}_4(\frac{h}{2})\circ
\tilde{\mathcal{H}}_3(\frac{h}{2})\circ
\tilde{\mathcal{H}}_2(\frac{h}{2}) \nonumber
\\ && \circ \tilde{\mathcal{H}}_1(h)\circ \tilde{\mathcal{H}}_2(\frac{h}{2})\circ
\tilde{\mathcal{H}}_3(\frac{h}{2})\circ
\tilde{\mathcal{H}}_4(\frac{h}{2}) \nonumber
\\ && \circ \tilde{\mathcal{H}}_5(\frac{h}{2}).
\end{eqnarray}
Its detailed expression is given in Appendix B.

A symmetric composition of three second-order methods yields a
fourth-order explicit symplectic scheme of Yoshida (1990)
\begin{equation}
S^{\mathcal{H}}_4(h)=S^{\mathcal{H}}_2(\gamma h)\circ
S^{\mathcal{H}}_2(\delta h)\circ S^{\mathcal{H}}_2(\gamma h),
\end{equation}
where $\delta=1-2\gamma$ and $\gamma=1/(2-\sqrt[3]{2})$.

Seen from the construction of the two explicit symplectic
integrators for the Hamiltonian $\mathcal{H}$, the time
transformation function $g$ plays an important role in
successfully eliminating $\Sigma$ in the denominators of the $p_r$
and $p_{\theta}$ terms of the inseparable Hamiltonian $H$. This is
successful to overcome an obstacle to the application of such
explicit symplectic integrators to the Hamiltonian $\mathcal{H}$.
Because $g=1+a^2\cos^{2}\theta/r^2\leq 1+a^2/r^2\leq
1+1/r^2\approx 1$, we have $\Delta\tau\sim g\Delta w\approx\Delta
w=h$ in Equation (23).\footnote{The relation between the proper
time step $\Delta\tau$ and the new coordinate time step $h$ does
not resemble that in Equation (4). The relation is accurately
given in Equation (4). Equations (33)-(37) are the differential
equations with respect to the new coordinate $w$, but Equation (4)
gives the evolution of the Keplerian motion in the physical time.
Because the evolution of the variables with the proper time is not
necessarily known in the present problem, a more accurate
description of the relation is not, either.} When the fixed time
step $h$ is given to the coordinate time $w$, the proper time step
$\Delta\tau$ slightly depends on the radial distance $r$, and even
is almost the same fixed coordinate time step $h$. The symplectic
structure of the time-transformed Hamiltonian flow $\mathcal{H}$
is preserved for the fixed coordinate time step $\Delta w$, but is
not for the slightly variable proper time step $\Delta\tau$. A
large difference between the two sets of time steps may exist when
the time transformation function $g$ is altered. Other choices of
time transformation function are given in Appendix C.

In the above-mentioned three time transformation functions, the
time transformation functions $g_1$ and $g_2$ play important roles
in step-size selection procedures with adaptive proper time steps
when an invariant coordinate time step is adopted. The choice of
$g_1$ exerts a larger influence on the adjustment of proper time
steps than that of $g$, but brings a smaller influence than that
of $g_2$. Here,  we are mainly interested in applying a time
transformation to implement the construction of explicit
symplectic integrators for the Kerr spacetime geometry.
Considering this purpose, we take the time transformation function
$g$ in the following discussions.

\begin{table*}[htbp]
\centering \caption{Computational cost [i.e. CPU times (minute$'$
second$''$)] for the algorithms in Figs. 1 and 2.} \label{Tab1}
\begin{tabular}{lcccccccccccc}
\hline Method   & RK4   & EP2 &  EP2*  & IE2  & EP4  & IE4  & IE4* & S2 & S4 & S4*\\
\hline Fig. 1   & $2'4''$ & $1'33''$ & $15'45''$ & $1'53''$ & $3'37''$ & $5'48''$  &  $1'33''$ & / & / & / \\
\hline Fig. 2   & $1'51''$ & $1'21''$ & $13'43''$ & $1'43''$ & $3'8''$ & $5'26''$  & $1'27''$ & $57''$ & $2'50''$ & $43''$ \\
\hline
\end{tabular}
\end{table*}

\section{Numerical evaluations}

Compared with the newly proposed integrators, several existing
numerical methods are considered. They are RK4, fourth-order
implicit and explicit mixed symplectic algorithm (IE4), and
fourth-order extended phase-space explicit symplectic-like method
(EP4).

\subsection{Integrating the original Hamiltonian $H$ with the existing algorithms}

The sum of the second and third terms in Equation (18) is labeled
as $K$. It is solved in terms of the second-order implicit
midpoint rule (Feng 1986). Its corresponding operator is $IM2(h)$,
where $h$ is a proper time step. $F$ in  Equation (18) is  easily
solvable and corresponds to an operator $\psi^{F}_{h}$. The two
operators symmetrically compose a second-order implicit and
explicit mixed symplectic integrator for the original Hamiltonian
$H$ in Equation (18)
\begin{equation}
IE2(h)=\psi^{K}_{h/2}\circ IM2(h)\circ\psi^{K}_{h/2}.
\end{equation}
The computational efficiency of the method IE2 acting on $H$ is
superior to that of the algorithm IM2 acting on $H$. More details
on the implicit and explicit mixed symplectic methods were given
in the references (Lubich et al. 2010; Zhong et al. 2010; Mei et
al. 2013a, 2013b). Similar to the algorithm $S^{\mathcal{H}}_4(h)$
in Equation (44), a fourth-order implicit and explicit mixed
symplectic algorithm is expressed as
\begin{equation}
IE4=IE2(\gamma h)\circ IE2(\delta h)\circ IE2(\gamma h).
\end{equation}

On the other hand, the four-dimensional phase-space variables
$(r,\theta, p_r, p_{\theta})$ of $H$ is extended to
eight-dimensional phase-space variables
$(r,\theta,\tilde{r},\tilde{\theta}, p_r,$
$p_{\theta},\tilde{p}_r, \tilde{p}_{\theta})$ in a new Hamiltonian
\begin{equation}
\mathbb{H}=\mathbb{H}_1(r,\theta, \tilde{p}_r,
\tilde{p}_{\theta})+\mathbb{H}_2(\tilde{r},\tilde{\theta}, p_r,
p_{\theta}),
\end{equation}
where $\mathbb{H}_1(r,\theta, \tilde{p}_r,
\tilde{p}_{\theta})=\mathbb{H}_2(\tilde{r},\tilde{\theta}, p_r,
p_{\theta})= H(r,\theta, p_r, p_{\theta})$. $\mathbb{H}_1$ and
$\mathbb{H}_2$ are independently solved in analytical methods.
They correspond to operators $\mathcal{A}$ and $\mathcal{B}$
similar to those in Sect. 2. The second- and fourth-order methods
$\mathcal{S}^{\mathbb{H}}_2$ and $\mathcal{S}^{\mathbb{H}}_4$ like
those in Equations (5) and (6) can be obtained. For the same
initial conditions, the two independent Hamiltonians should have
the same solutions: $r=\tilde{r}$, $\theta=\tilde{\theta}$,
$\tilde{p}_r= p_r$ and $\tilde{p}_{\theta}=p_{\theta}$.  However,
the two solutions are not the same due to their couplings in
$\mathcal{S}^{\mathbb{H}}_2$ and $\mathcal{S}^{\mathbb{H}}_4$.
Permutations between the original variables and their
corresponding extended variables are necessarily used after the
implementation of $\mathcal{S}^{\mathbb{H}}_2$ and
$\mathcal{S}^{\mathbb{H}}_4$ so as to frequently make the two
solutions  equal (Pihajoki 2015; Liu et al. 2016; Luo et al. 2017;
Liu et al. 2017; Luo $\&$ Wu 2017; Li $\&$ Wu 2017; Wu $\&$ Wu
2018). The midpoint permutation method (Luo et al. 2017)
\begin{eqnarray}
\mathcal{M}: \frac{r+\tilde{r}}{2} &\rightarrow& r
=\tilde{r},\nonumber \\
 \frac{\theta+\tilde{\theta}}{2} &\rightarrow&
\theta=\tilde{\theta};
\nonumber \\
\frac{p_r+\tilde{p}_r}{2} &\rightarrow& p_r=\tilde{p}_r,  \\
\frac{p_{\theta}+\tilde{p}_{\theta}}{2} &\rightarrow& p_{\theta}=
\tilde{p}_{\theta} \nonumber
\end{eqnarray}
is a good choice. The algorithms $\mathcal{S}^{\mathbb{H}}_2$ and
$\mathcal{S}^{\mathbb{H}}_4$ combined with the midpoint
permutation are
\begin{eqnarray}
EP2 &=& \mathcal{M}\otimes \mathcal{S}^{\mathbb{H}}_2, \\
EP4 &=& \mathcal{M}\otimes \mathcal{S}^{\mathbb{H}}_4.
\end{eqnarray}
The symplecticity of $\mathcal{S}^{\mathbb{H}}_2$ and
$\mathcal{S}^{\mathbb{H}}_4$ is destroyed due to the inclusion of
$\mathcal{M}$. However, the symmetry makes the methods IE2 and IE4
exhibit good long-term stable behavior in Hamiltonian errors. In
this sense, the methods IE2 and IE4 are explicit symplectic-like
algorithms for the newly extended phase-space Hamiltonian
$\mathbb{H}$ in Equation (47).

Taking the parameters $E=0.995$, $L=4.6$ and $a=0.5$, we choose a
test orbit with the initial conditions $r=11$, $\theta=\pi/2$ and
$p_r=0$. The starting value of $p_{\theta}>0$ is given by Equation
(17). The proper time step is $h=1$ and the integration time
arrives at $10^{8}$. As shown in Figure 1, RK4 shows a secular
drift in Hamiltonian error $\Delta H=-1-2H$ in Equation (20). So
does the second-order extended phase-space method EP2. However,
this drift is absent in EP2* when the proper time step $h=1$ is
replaced with $h=0.1$. It is because the extended phase-space
method maintaining the boundness of errors requires appropriate
permutations. When the step-size becomes smaller, more
permutations are used. This results in the improvement of the
algorithm's accuracy. In fact, the accuracy of EP2* has an order
of $\mathcal{O}(10^{-8})$ and is two orders of magnitude higher
than that of EP2. The fourth-order extended phase-space method EP4
with the proper time step $h=1$ also gives the same order errors
and therefore makes the errors bounded. If the Hamiltonian
truncation errors are not larger than the order of
$\mathcal{O}(10^{-9})$, $10^{8}$ integration steps yield a great
number of roundoff errors that cause the Hamiltonian errors to
have a secular growth.\footnote{For a computer with double
precision $\epsilon$, per calculation  may yield a roundoff error
$\epsilon$. The roundoff errors after $n=10^{8}$ integration steps
are roughly estimated to be $n\epsilon\approx 10^{-8}$. When the
truncation error of an algorithm is smaller than the order of
$\mathcal{O}(10^{-8})$, the roundoff errors have an important
influence on the truncation error.} The fourth-order implicit and
explicit mixed symplectic algorithm IE4 is suitable for this case.
When a larger step-size $h=4$ is used, the Hamiltonian errors for
IE4* are similar to those for the second-order implicit and
explicit mixed symplectic algorithm IE2 with the proper time step
$h=1$ and remain stable and bounded.

In short, the algorithms' performance tested in the above Kerr
geometry is very similar to that checked in the Schwarzschild
spacetime in Paper I (Wang et al. 2021a).

\subsection{Solving time-transformed Hamiltonian $\mathcal{H}$}

Without question, the algorithms RK4, IE4 and EP4 are also suited
for solving the time-transformed Hamiltonian $\mathcal{H}$ in
Equation (26). In addition to them, the new explicit symplectic
methods $S^{\mathcal{H}}_2$ (labeled as S2) and
$S^{\mathcal{H}}_4$ (labeled as S4) are respectively used to solve
the Hamiltonian splitting form (27). Now, $h$ is a coordinate time
step. It is shown by the comparison between Figure 1 and Figures.
2 (a) and (b) that the existing algorithms for the
time-transformed Hamiltonian $\mathcal{H}$ and those for the
original Hamiltonian $H$ in Equation (18) have no typical
differences in the Hamiltonian error behavior. The methods S2 and
S4 exhibit good long-term stabilized error behavior in Figure 2
(c). No secular error change exists for S4 because S4 gives  the
order of $\mathcal{O}(10^{-8})$ to the Hamiltonian errors rather
than the order of $\mathcal{O}(10^{-9})$.

The results can be concluded from Figure 2. For the coordinate
time step $h=1$, the algorithms RK4, EP2, IE2 and S2 are
approximately the same in the Hamiltonian errors; the fourth-order
methods EP4, IE4 and S4 are, too. The schemes without secular
error drifts are IE2, S2, EP4 and S4. In addition, EP2* with a
smaller coordinate time step $h=0.1$ approaches to S4 in
accuracy. IE4* and S4* with  a larger coordinate time step $h=4$
are close to S2. EP2*, IE* and S4* maintain the boundness of
Hamiltonian errors. These similar results are still present when
the dynamical parameters and orbits are altered.

Besides Hamiltonian $\mathcal{H}$, Carter constant $K$ in Appendix
A can be conserved by the algorithms S4, EP4 and IE4, but cannot
be conserved by the RK4 method, as shown in Figure 3 (a). RK4 is
the poorest in accuracy of the Carter constant, and EP4 is the
best.  IE4 is slightly better than S4. However, RK4 is the best in
accuracy of the solution for an integration time of $w=2000$ in
Figure 3 (b), and S4 is better than EP4 or IE4. This shows that
RK4 is accurate and efficient for such a short time integration.

Testing the performance of these algorithms in Figures 1-3 is
based on the boundness of  massive particle orbits. What about the
performance of these algorithms if the integrated orbits are
unstable and unbounded? To answer this question, we adopt the
unstable spherical photon orbit B with normalized angular momentum
$\Phi=-6$, normalized Carter constant $Q=-13+16\sqrt{2}$ and
constant radius $r_0=1+2\sqrt{2}$ considered by Bacchini et al.
(2018). Some details of the unstable spherical photon orbits are
described in Appendix A. When $w=116$ in Figure 4, the orbit
begins to run to infinity for RK4, and accuracies of Hamiltonian
$\mathcal{H}$ and Carter constant $K$ become the worst. Although
this orbit does not remain spherical due to orbital instability,
S4, IE4 and EP4 can still work well in conservation of the two
constants. In fact, S4 exhibits the best accuracy.

Apart from accuracies of the constants and solutions for these
algorithms, computational efficiency should be compared. Table 1
lists CPU times of the algorithms in Figures 1 and 2.  S2 has the
best efficiency among the three second-order methods S2, IE2 and
EP2. The efficiency of S4 is also the best among the efficiencies
of the three fourth-order methods S4, IE4 and EP4. This shows the
superiority of the application of explicit algorithms in
computational efficiency.

Figure 5 (a) plots the evolution curve  of $r$ with proper time
$\tau$ (colored black dot), obtained by IE4 solving the original
Hamiltonian system (18) with the proper time step $h=1$. It also
plots the evolution of $r$ with coordinate time $w$ (colored red
line), given by  S4 integrating the time-transformed Hamiltonian
(26) with the coordinate time step $h=1$. The two evolution curves
coincide. Here is an explanation. The proper time steps are varied
for the use of fixed coordinate time step. However, the coordinate
time $w$ and the proper time $\tau$ have no typical differences
when S4 solves the Hamiltonian (26) with the coordinate time step,
as shown in Figure 5 (b). Clearly, the time transformation
function in Equation (25) mainly plays an important role in
eliminating the function $\Sigma$ in the denominators of the
Hamiltonian (18). Then it leads to successfully implementing the
construction of explicit symplectic integrators for the
time-transformed Hamiltonian (26). The time transformation
function does not have any explicit effect on step-size selection
procedures.

\subsection{Discussions}

The new explicit symplectic integrators are applicable for not
only integrating geodesics in the Kerr spacetime, but also the
Kerr black hole immersed in an external magnetic field
(Kop\'{a}\v{c}ek et al. 2010; Kop\'{a}\v{c}ek $\&$ Karas 2014;
Kop\'{a}\v{c}ek $\&$ Karas 2018), Kerr-Newman black hole, and
Kerr-Newman black hole with an external magnetic field. The
construction of explicit symplectic integrators is based on the
splitting of the time-transformed Hamiltonian when the
denominators  of contravariant metric components $g^{rr}$ and
$g^{\theta\theta}$ have $\Sigma$ as a function  of $r$ and
$\theta$, and the time transformation function $g=\Sigma/r^2$ is
taken. Such similar constructions are still possibly available if
$g^{rr}$ and $g^{\theta\theta}$ have other expressions. In this
case, the time transformation function $g$ should need an
appropriate choice. In fact, the splitting is valid if the
time-transformed Hamiltonian becomes
\begin{eqnarray}
\mathcal{H} &=&
K_1(r,\theta)+gg^{rr}p^2_r+gg^{\theta\theta}p^2_{\theta\theta},  \\
gg^{rr}&=& (a_0+a_1r+\cdots+a_nr^n+\frac{a_{-1}}{r}+\frac{a_{-2}}{r^2} \nonumber \\
&& +\cdots+\frac{a_{-m}}{r^m}) K_2(\theta),  \\
gg^{\theta\theta} &=& (b_0+b_1\theta+\cdots+b_i\theta^i+\frac{b_{-1}}{\theta}+\frac{b_{-2}}{\theta^2} \nonumber \\
&& +\cdots+\frac{b_{-j}}{r^j}) K_3(r),
\end{eqnarray}
where $K_1$,  $K_2$ and  $K_3$ are arbitrary continuous
differentiable functions when $r$ and $\theta$ are restricted to
$r>2$ and $\theta\neq0$, and $a_0$, $\cdots$,  $a_n$, $a_{-1}$,
$\cdots$, $a_{-m}$,  $b_0$, $\cdots$,  $b_i$, $b_{-1}$, $\cdots$,
$b_{-j}$ are constant parameters. It is clear that the
time-transformed Hamiltonian splitting is desired when time
transformation function $g$ satisfies Equations (52) and (53). For
example, the time transformation function takes $g=e^{Q-P}$ for
$g^{rr}=e^{P-Q}(1-2/r)$ and $g^{\theta\theta}=e^{P-Q}r^{-2}$,
where $P$ and $Q$ are  functions of $r$ and $\theta$  in an
axially symmetric core-shell system (Vieira $\&$ Letelier 1999).
Another example is a five-dimensional black ring metric (Igata et
al. 2011), where $g^{xx}=(x-y)^2(1-x^2)(1+\nu x)/[R^2(1+\lambda
x)]$ and $g^{yy}=-(x-y)^2(1-y^2)(1+\nu y)/[R^2(1+\lambda x)]$ in
ring coordinates $(t, x, y, \phi,\psi)$. Here, $|x|\leq1$ and
$y\leq-1$ are coordinate variables; $R>0$ and $0<\nu\leq\lambda<1$
are constant parameters; $t$, $\phi$ and $\psi$ do not explicitly
appear in the metric. Obviously, time transformation function
$g=1+\lambda x$ is a good choice. Therefore, the idea for the
construction of the present explicit symplectic integrators is not
restricted to the Kerr type spacetimes, and is applied to many
other relativistic problems.

Although the application of the new algorithms is not wider than
that of RAPTOR based on the RK4 scheme (Bronzwaer et al. 2018),
accuracies of the new algorithms have an advantage over those of
the RK4 scheme in long-term integrations. The new second-order
explicit symplectic integrator S2 and RK4 have the same order of
magnitude in the Hamiltonian errors in Figures 2 (a) and (c), but
they have typical differences in simulations of unstable spherical
photon orbits or long-term evolution of massive particle bounded
orbits. The errors grow with an increase of integration time for
RK4, but remain bounded for S2. When the integration time reaches
116, RK4 gives extremely larger errors to the constants for the
unstable spherical photon orbit, but S4 shows smaller errors. When
the integration time lasts $10^{10}$, RK4 does not work well and
provides uncorrect results to massive particle bounded orbits,
whereas S2 still shows no secular growth in the errors. This is an
advantage of a symplectic scheme in long-term integrations. In
particular, Deng et al. (2020) reported that RK4 combined with
manifold correction exhibits poorer performance than the
second-order symplectic leapfrog  in a $10^8$ year integration of
the outer solar system involving the Sun and four outer planets.
In addition, Figures 2 (a) and (c) clearly show that the new
fourth-order explicit symplectic integrator S4 is two orders of
magnitude higher than RK4 in accuracy. S4 is typically better than
RK4 in numerical performance for the integration of unstable
spherical photon orbits.

As shown in Table, the newly proposed algorithms are greatly
superior to the implicit and explicit mixed symplectic methods
without doubt in terms of computational efficiency. They have also
an advantage over the extended phase-space explicit
symplectic-like methods EP2 and EP4 because they integrate over 4
dimensions, while the extended phase-space schemes require
doubling the phase space (i.e., 8 dimensions). Figures 2 (a) and
(c) describe that the extended phase-space method EP2 possesses
larger errors than the proposed algorithm S2. Figure 3 (b) shows
that EP4 is poorer than S4 in accuracy of the solutions. The
methods EP2 and EP4 are in fact similar to the method of Tao
(2016) or FANTASY (Christian $\&$ Chan 2021). Wu $\&$ Wu (2018)
found that the fourth-order methods like EP4 with the midpoint
permutations show better accuracies than the Tao's method or
FANTASY with  a good choice of the control constant.

The newly proposed algorithms are inferior to the
energy-conserving discrete gradient scheme of Bacchini et al.
(2018) in energy conservation, but have some explicit advantages
over the latter method. The latter algorithm are implicit,
nonsymplectic, and do not preserve other integrals in general
(perhaps some integrals may be conserved, as claimed in Footnote
1), but the proposed algorithms are explicit, symplectic, and
preserve other integrals like the Carter constant. Therefore, the
proposed algorithms are extremely superior to the latter algorithm
in computational efficiency. In addition, fourth, sixth orders and
higher order schemes are easily available according to the present
idea, whereas they are not in terms of the construction method of
Bacchini et al.

\section{Conclusions}

The function $\Sigma$ depending on $r$ and $\theta$ exists in the
denominators of the Hamiltonian of Kerr geometry. Thus, explicit
symplectic integrators are useless if the Hamiltonian is split
into several parts like  the Hamiltonian of Schwarzschild
spacetimes in the previous papers (Wang et al. 2021a, 2021b,
2021c).

To overcome this difficulty, we use an appropriate time
transformation function to eliminate the function $\Sigma$ in the
denominators and to obtain a time-transformed Hamiltonian. This
Hamiltonian can be separated into five parts with analytical
solutions as explicit functions of the new coordinate time, and
therefore explicit symplectic integrators are easily available.
Numerical tests show that the explicit symplectic methods perform
good long-term stable behavior in Hamiltonian errors when
appropriate coordinate time steps are chosen. The use of fixed
coordinate time steps maintains the symplecticity of the
time-transformed Hamiltonian. Although the proper time sizes are
variant, there is not an explicit difference between the proper
time and newly transformed coordinate time. In this sense,  the
chosen time transformation function does not mainly play a role in
step size selection procedures. The proposed algorithms are
superior to the same order extended phase-space explicit
symplectic-like methods and implicit and explicit mixed symplectic
algorithms in computational efficiency.

The new idea for the construction of such explicit symplectic
integrators is not limited to the application of the Kerr metric,
and allow for the application of many spacetimes given in Equation
(51). This problem will further be considered in our future works.

\appendix
%\section*{APPENDIX}

\section{Carter constant and unstable spherical
photon orbits}

Noticing Equations (18)-(20), we rewrite Equation (18) as follows:
\begin{equation}
-\chi\Sigma = -\frac{E^2}{\Delta}(\rho^4-\Delta a^2\sin^2\theta)
+\frac{L^2(\Delta-a^2\sin^{2}\theta)}{\Delta \sin^{2}\theta}
+\frac{4ar}{\Delta}EL +\Delta p^{2}_{r} +p^{2}_{\theta},
\end{equation}
where $\chi=1$ for massive particles and $\chi=0$ for massless
photons.  The equation has the separation of variables:
\begin{equation}
p^{2}_{\theta}+\frac{L^2}{\sin^{2}\theta}+E^2a^2\sin^2\theta+\chi
a^2\cos^2\theta =-\chi r^2
+\frac{E^2}{\Delta}\rho^4+\frac{L^2a^2}{\Delta}
-\frac{4ar}{\Delta}EL -\Delta p^{2}_{r}.
\end{equation}
Clearly, the left-hand side is a function of variables $\theta$
and $p_{\theta}$, whereas  the right-hand side is a function of
variables $r$ and $p_{r}$. This equality is possible only when the
two sides are equal to same constant
\begin{eqnarray}
K &=& p^{2}_{\theta}+[\frac{L^2}{\sin^{2}\theta}
+a^2(\chi-E^2)]\cos^2\theta, \\
K &=& -L^2-a^2E^2-\chi r^2
+\frac{E^2}{\Delta}\rho^4+\frac{L^2a^2}{\Delta}
-\frac{4ar}{\Delta}EL -\Delta p^{2}_{r}.
\end{eqnarray}
Equation (A3) or (A4) is an expressional form of the Carter
constant.

Equation (A4) is rewritten as
\begin{eqnarray}
\Sigma^2 (\frac{dr}{d\tau})^2 = -(K+L^2+a^2E^2+\chi r^2)\Delta
+E^2\rho^4+L^2a^2 -4arEL=R(r).
\end{eqnarray}
When $R(r)=dR(r)/dr=0$ in Equation (A5), $r$ corresponds to a
constant radius of spherical photon orbit. Teo (2003) studied such
spherical photon orbits with normalized angular momentum $\Phi$
and  normalized Carter constant $Q$:
\begin{eqnarray}
\Phi &=& \frac{L}{E} =-\frac{r^3-3r^2+a^2(r+1)}{a(r-1)}, \\
Q &=& \frac{K}{E^2}=-\frac{r^3(r^3-6r^2+9r-4a^2)} {[a(r-1)]^2}.
\end{eqnarray}
These two expressions were also provided by Chan et al. (2018).
Due to $d^2R(r)/dr^2>0$, these orbits would be unstable when they
are suffered from perturbations in the radial direction.
M\"{u}ller $\&$ Grave (2010) gave the expression of $E$ as
follows:
\begin{eqnarray}
E=\left(\sqrt{\frac{A}{\Sigma\Delta}}-\frac{2ra\Phi}{\sqrt{A\Sigma\Delta}}\right)^{-1}.
\end{eqnarray}
Thus, the other two parameters are easily determined by
\begin{eqnarray}
L=E\Phi, ~~~~~~~~ K=QE^2.
\end{eqnarray}
For given parameters  $r$ and $a=1$ with the initial value
$\theta=\pi/2$, we use Equation (A3) to obtain the initial value
\begin{eqnarray}
p_{\theta}=E\sqrt{Q}.
\end{eqnarray}

\section{Implementation of algorithm $S^{\mathcal{H}}_2$}

Codes of the method in Equation (43) are written in the following
forms
\begin{eqnarray}
\theta^{\ast} &=& \theta_0+\frac{h}{2}\frac{p_{\theta0}}{r^2_0}, \nonumber \\
p^{\ast}_{r} &=& p_{r0}+\frac{h}{2}\frac{p^2_{\theta0}}{r^{3}_0}; \nonumber \\
r^{\star} &=& (\frac{h}{r_0}a^{2}p^{\ast}_{r}+r^2_0)^{1/2}, \nonumber \\
 p^{\star}_{r} &=& \frac{p^{\ast}_{r}}{r_0}(\frac{h}{r_0}a^{2}p^{\ast}_{r}+r^2_0)^{1/2}; \nonumber \\
r^{\dag} &=& [(r^{\star2}-\frac{3}{2}h
p^{\star}_{r})^{2}/r^{\star}]^{1/3}, \nonumber \\
 p^{\dag}_r &=& p^{\star}_{r}[(r^{\star2}-\frac{3}{2}h
p^{\star}_{r})/r^{\star2}]^{1/3}; \nonumber \\
r^{\ddag} &=& r^{\dag}+\frac{h}{2} p^{\dag}_r; \nonumber \\
p^{\sharp}_r &=& p^{\dag}_r+h P_{r}(r^{\ddag},\theta^{\ast}), \nonumber \\
\tau_1 &=& \tau_0+ h g(r^{\ddag},\theta^{\ast}), \nonumber \\
p_{\theta1} &=& p_{\theta0}+h P_{\theta}(r^{\ddag},\theta^{\ast}); \nonumber \\
r^{\natural} &=& r^{\ddag}+\frac{h}{2} p^{\sharp}_r; \nonumber \\
r^{\diamondsuit} &=& [(r^{\natural2}-\frac{3}{2}h
p^{\sharp}_{r})^{2}/r^{\natural}]^{1/3}, \nonumber \\
 p^{\diamondsuit}_r &=& p^{\sharp}_{r}[(r^{\natural2}-\frac{3}{2}h
p^{\sharp}_{r})^{2}/r^{\natural2}]^{1/3}; \nonumber \\
 p^{\heartsuit}_{r} &=& \frac{p^{\diamondsuit}_{r}}{r^{\diamondsuit}}(\frac{h}{r^{\diamondsuit}}a^{2}
 p^{\diamondsuit}_{r}+r^{\diamondsuit2})^{1/2}, \nonumber \\
 r_{1} &=& (\frac{h}{r^{\diamondsuit}}a^{2}p^{\diamondsuit}_{r}+r^{\diamondsuit2})^{1/2}; \nonumber \\
p_{r1} &=& p^{\heartsuit}_{r}+\frac{h}{2}\frac{p^2_{\theta1}}{r^{3}_1}, \nonumber \\
\theta_{1} &=& \theta^{\ast}+\frac{h}{2}\frac{p_{\theta1}}{r^2_1}.
\nonumber
\end{eqnarray}

In this way, the solutions
$(\tau_1,r_1,\theta_1,p_{r1},p_{\theta1})$  are outputted after
the values $(\tau_0,r_0,\theta_0,p_{r0},p_{\theta0})$  advance a
fixed coordinate time step $h$.

\section{Other choices of time transformation function}

In principle,  the time transformation function has various
choices. For instance, it is given by
\begin{equation}
g_{1}=\frac{\Sigma}{r}=r+\frac{1}{r}a^2\cos^{2}\theta.
\end{equation}
The new coordinate time in Equation (23) becomes $w_1$. Advancing
the fixed coordinate time $\Delta w_1=h$ means advancing a
variable proper time $\Delta \tau_1\approx rh$. The advancement of
proper time  is absolutely dominated by the radial separation $r$.
When a particle approaches to the central black hole\footnote{For
example, the presence of close encounters between the black hole
and the particle moving in a highly eccentric orbit belongs to
this case.}, $\Delta \tau_1$ is smaller; but it becomes larger
when the particle moves always from the central object. This kind
of step-size selection procedure uses adaptive time steps, and
satisfies the  realistic need on the improvement of numerical
precisions. An important contribution for the use of time
transformation equation  is that this transformation brings the
implementation of symplectic integrators with adaptive time step
controls. This is the main result claimed in the paper of Mikkola
(1997). In terms of the time transformation function $g_{1}$,
another new time-transformed Hamiltonian has four separable parts
\begin{equation}
\mathcal{H}^{\ast}=\mathcal{H}^{\ast}_1+\mathcal{H}^{\ast}_2+\mathcal{H}^{\ast}_3+\mathcal{H}^{\ast}_4,
\end{equation}
where these parts are
\begin{eqnarray}
\mathcal{H}^{\ast}_1 &=& \frac{\Sigma}{r} (F+p_0), \\
\mathcal{H}^{\ast}_{2} &=& (\frac{r}{2}-1)p^{2}_{r},\\
\mathcal{H}^{\ast}_{3} &=& \frac{a^2}{2r}p^{2}_{r},\\
\mathcal{H}^{\ast}_{4} &=& \frac{1}{2r}p^{2}_{\theta}.
\end{eqnarray}
Like the method $S^{\mathcal{H}}_2$, another new explicit
second-order symplectic integrator for $\mathcal{H}^{\ast}$ reads
\begin{equation}
S^{\mathcal{H}^{\ast}}_2(h) =
\tilde{\mathcal{H}}^{\ast}_4(\frac{h}{2})\circ
\tilde{\mathcal{H}}^{\ast}_3(\frac{h}{2})\circ
\tilde{\mathcal{H}}^{\ast}_2(\frac{h}{2})\circ
\tilde{\mathcal{H}}^{\ast}_1(h) \circ
\tilde{\mathcal{H}}^{\ast}_2(\frac{h}{2})\circ
\tilde{\mathcal{H}}^{\ast}_3(\frac{h}{2})\circ
\tilde{\mathcal{H}}^{\ast}_4(\frac{h}{2}).
\end{equation}
A fourth-order method $S^{\mathcal{H}^{\ast}}_4$ can also be
obtained. The new explicit symplectic algorithms in the new
coordinate time $w_1$ are similar to those in the proper time of
Schwarzschild spacetime geometry in Paper I (Wang et al. 2021a).

If the time transformation function takes the form
\begin{equation}
g_{2}(r,\theta)=\Sigma,
\end{equation}
the new coordinate time in Equation (23) is $w_2$. In this case,
Equation (24) corresponds to a simpler time-transformed
Hamiltonian
\begin{equation}
\mathcal{H}^{\star}=\Sigma (F+p_0) +\frac{\Delta}{2}p^{2}_{r}
+\frac{1}{2}p^{2}_{\theta}.
\end{equation}
It has three separable parts
\begin{equation}
\mathcal{H}^{\star}=\mathcal{H}^{\star}_1+\mathcal{H}^{\star}_2+\mathcal{H}^{\star}_3,
\end{equation}
where the sub-Hamiltonians are
\begin{eqnarray}
\mathcal{H}^{\star}_1 &=& \Sigma (F+p_0), \\
\mathcal{H}^{\star}_{2} &=& \frac{1}{2}r^2p^{2}_{r},\\
\mathcal{H}^{\star}_{3} &=&
\frac{1}{2}(a^2-2r)p^{2}_{r}+\frac{1}{2}p^{2}_{\theta}.
\end{eqnarray}
Obviously, the three parts are independently solved in analytical
methods and their analytical solutions are explicit functions of
the new coordinate time $w_2$. Using Lie operators
$\tilde{\mathcal{H}}^{\star}_{1}$,
$\tilde{\mathcal{H}}^{\star}_{2}$ and
$\tilde{\mathcal{H}}^{\star}_{3}$ to represent the analytical
solutions of $\mathcal{H}^{\star}_{1}$, $\mathcal{H}^{\star}_{2}$
and $\mathcal{H}^{\star}_{3}$, we easily obtain an explicit
second-order symplectic integrator for the system
$\mathcal{H}^{\star}$ as follows
\begin{equation}
S^{\mathcal{H}^{\star}}_2(h) =
\tilde{\mathcal{H}}^{\star}_3(\frac{h}{2})\circ
\tilde{\mathcal{H}}^{\star}_2(\frac{h}{2})\circ
\tilde{\mathcal{H}}^{\star}_1(h) \circ
\tilde{\mathcal{H}}^{\star}_2(\frac{h}{2})\circ
\tilde{\mathcal{H}}^{\star}_3(\frac{h}{2}).
\end{equation}
An explicit fourth-order symplectic integrator
$S^{\mathcal{H}^{\star}}_4$ can be easily available, too. As a
point to illustrate, the newly fixed coordinate time step $\Delta
w_2=h$ corresponds to a variant proper time step $\Delta
\tau_2\approx r^2h$.

For $r\gg1$, the fixed coordinate time steps $\Delta w_1\ll1$ and
$\Delta w_2\ll1$ should be required when the variant proper time
steps $\Delta \tau_1$ and $\Delta \tau_2$ are 1. Such extremely
small coordinate time steps would lead to fatal roundoff errors.
Thus, we take the time transformation function (25) rather than
Equations (C1) and (C8).

\section*{Acknowledgments}

The authors are very grateful to a referee for valuable comments
and useful suggestions. This research has been supported by the
National Natural Science Foundation of China [Grant Nos. 11973020
(C0035736), 11533004, 11803020, 41807437, U2031145], and the
Natural Science Foundation of Guangxi (Grant Nos.
2018GXNSFGA281007 and 2019JJD110006).

\newpage

\begin{figure*}[ptb]
\center{
\includegraphics[scale=0.25]{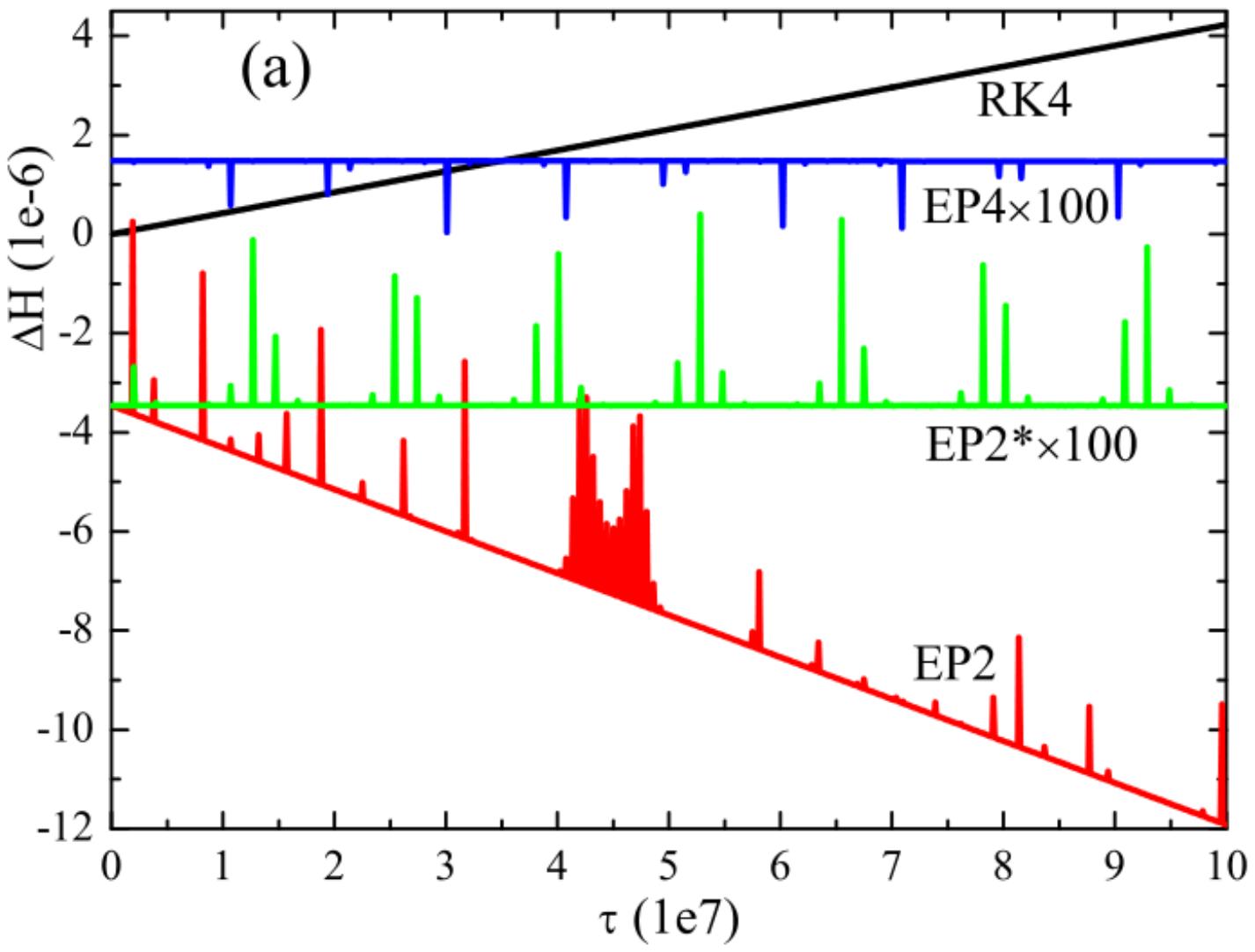}
\includegraphics[scale=0.25]{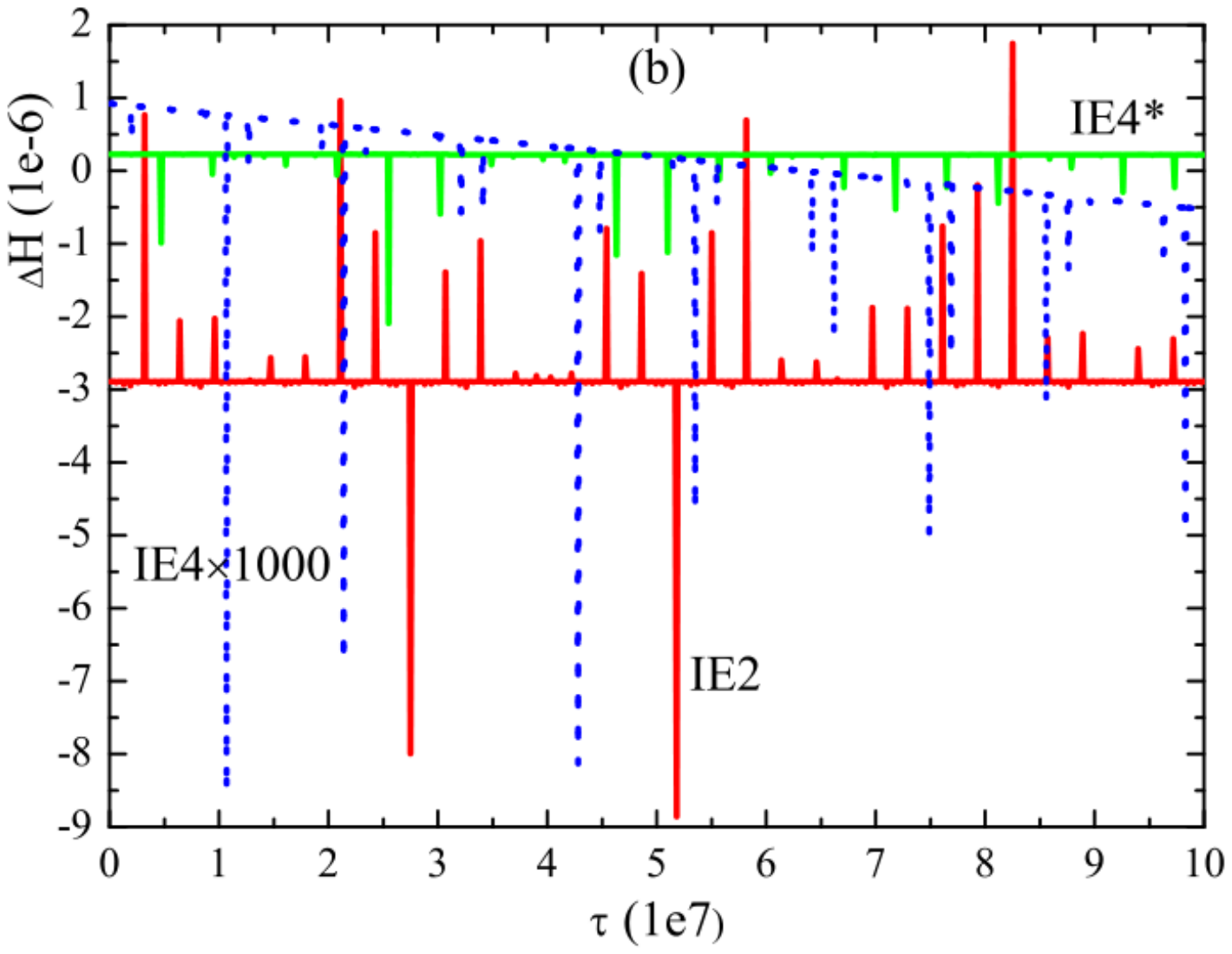}
\caption{Hamiltonian errors $\Delta H=-1-2H$ in Equation (20) when
several known numerical integration algorithms acting on the
original Hamiltonian system (18) for massive particles. The
algorithms include the second-order implicit and explicit mixed
symplectic method IE2, the second-order explicit extended
phase-space symplectic-like algorithm EP2, the fourth-order
implicit and explicit  mixed symplectic method IE4, the
fourth-order explicit extended phase-space symplectic-like
algorithm EP4 and the fourth-order Runge-Kutta scheme RK4. The
parameters are taken as $E=0.995$, $L=4.6$ and $a=0.5$. The
initial conditions of an orbit are $r=11$, $\theta=\pi/2$, $p_r=0$
and $p_{\theta}>0$ given by Equation (17). All integrations are
performed in the proper time. The proper time step is $h=0.1$ for
EP2*, $h=4$ for IE4*, and $h=1$ for the other methods. The
notation EP2*$\times$100 means that the plotted errors for EP2*
are enlarged 100 times, compared with the realistic errors. }
 \label{Fig1}}
\end{figure*}

\begin{figure*}[ptb]
\center{
\includegraphics[scale=0.18]{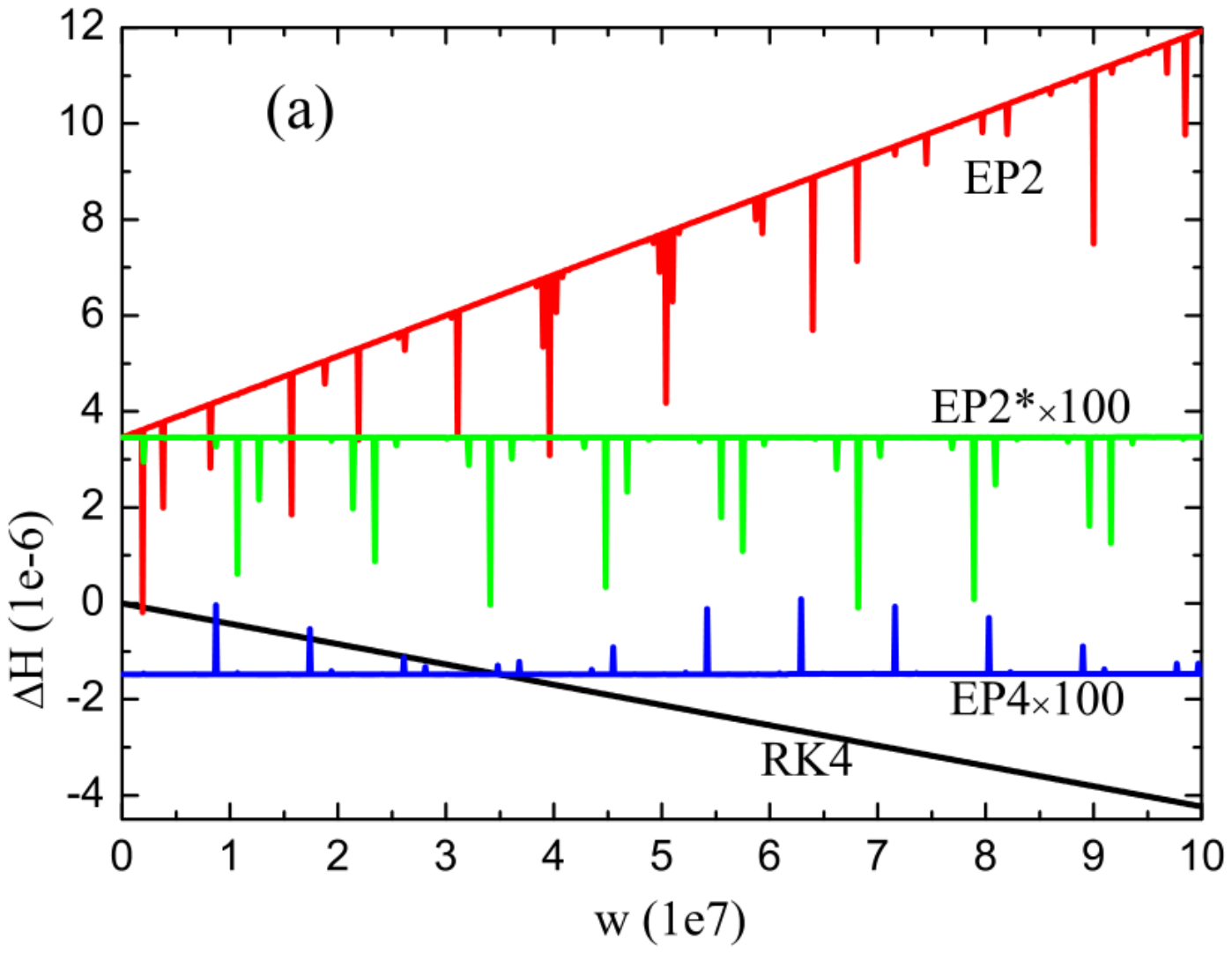}
\includegraphics[scale=0.18]{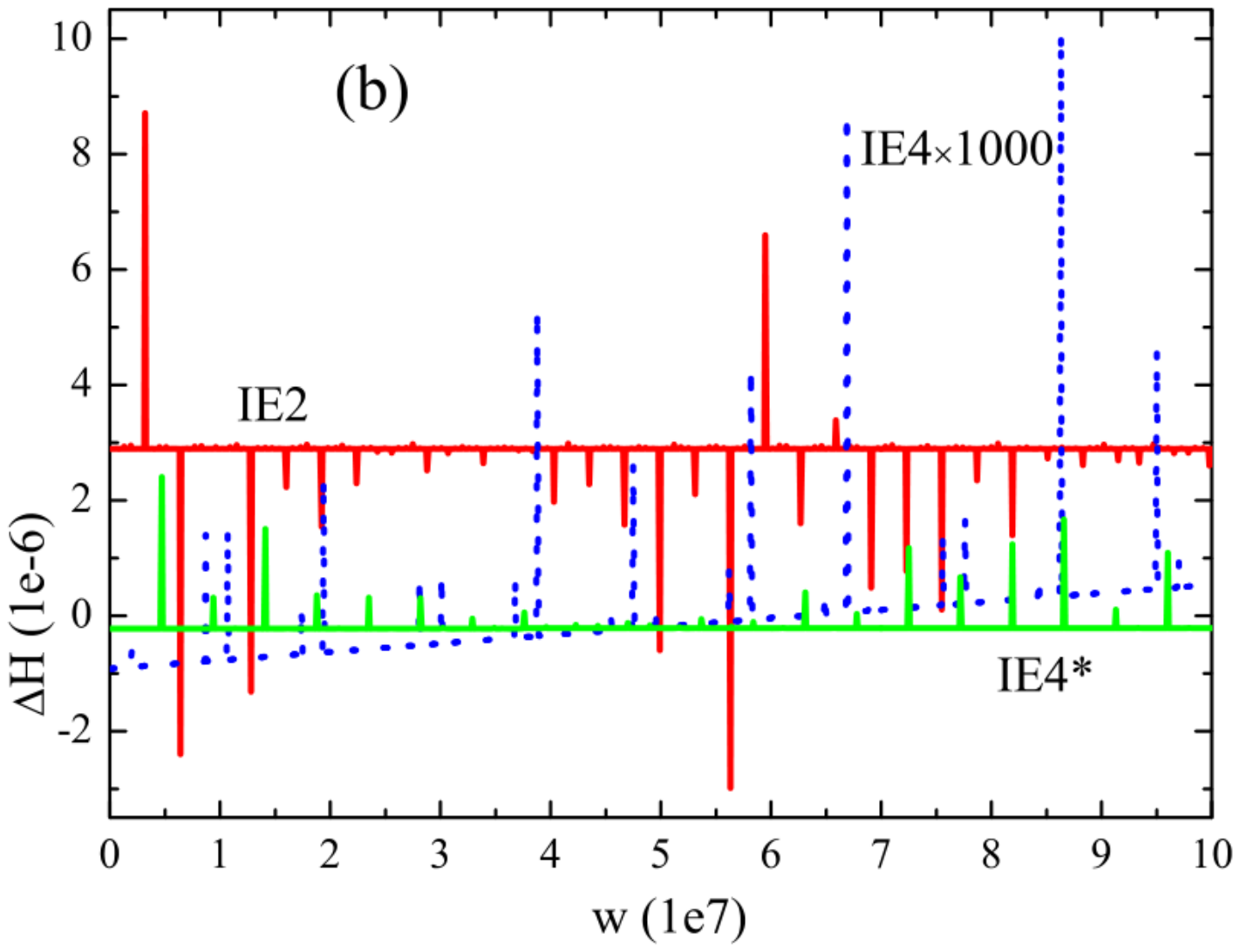}
\includegraphics[scale=0.18]{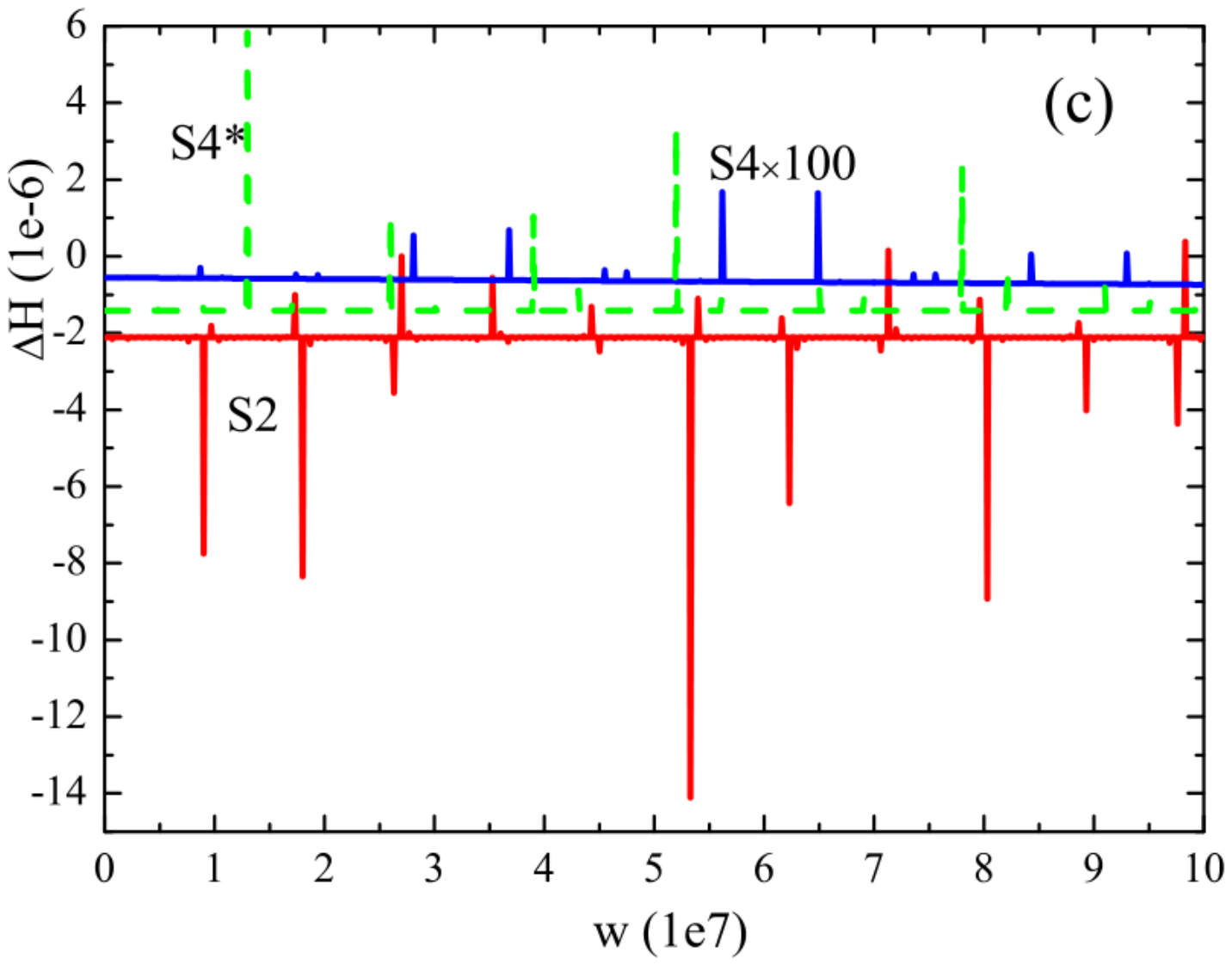}
\caption{Accuracy of the time-transformed Hamiltonian (26),
$\Delta H=2\mathcal{H}$. Panels (a) and (b) correspond to Figures
1 (a) and (b). Panel (c) relates to the new second- and
fourth-order explicit symplectic integrators S2 and S4. All
computations are implemented in the new coordinate time $w$. The
coordinate time step is $h=1$ for S2 and S4, and $h=4$ for S4*. }
 \label{Fig2}}
\end{figure*}

\begin{figure*}[ptb] \center{
\includegraphics[scale=0.25]{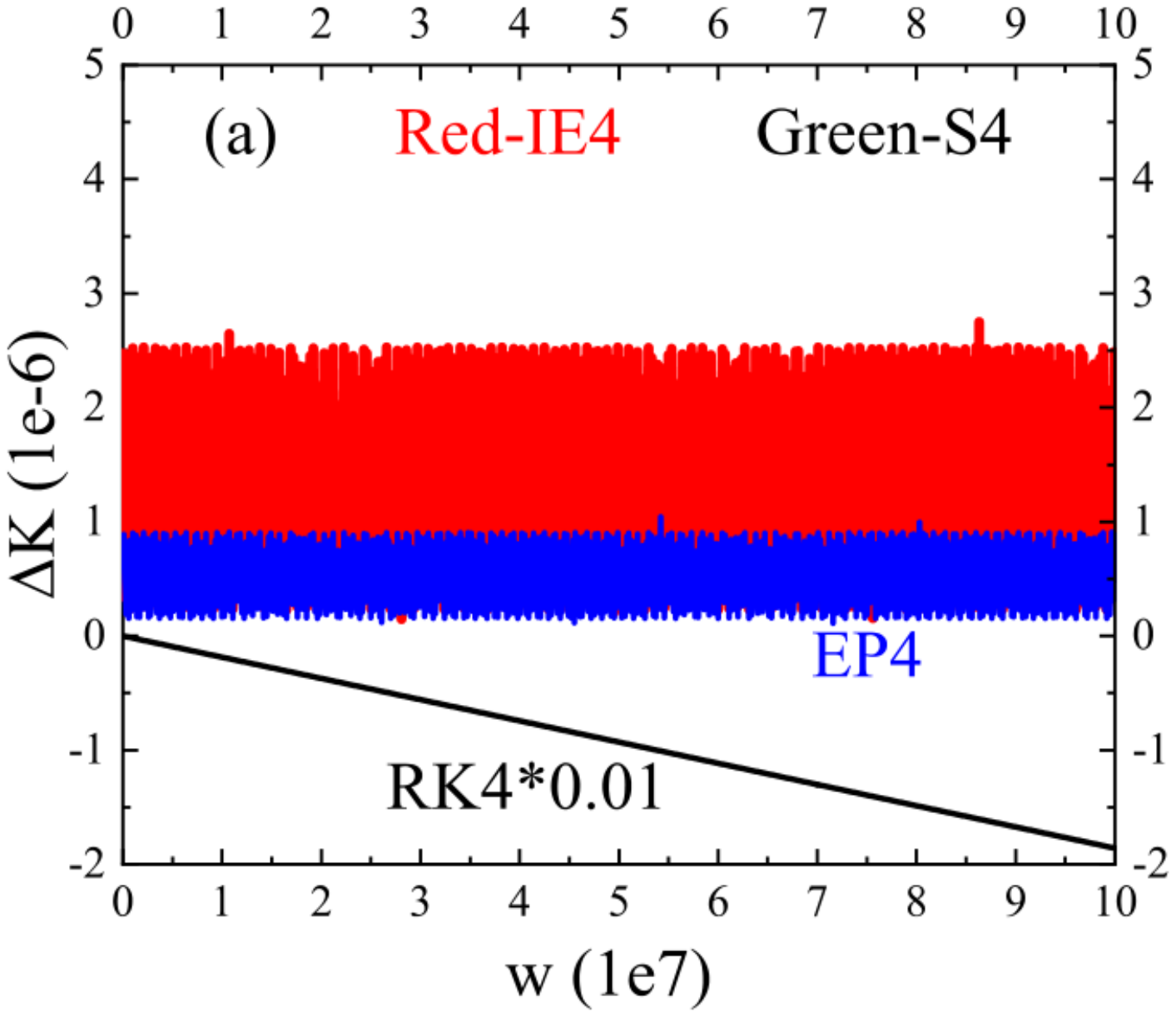}
\includegraphics[scale=0.25]{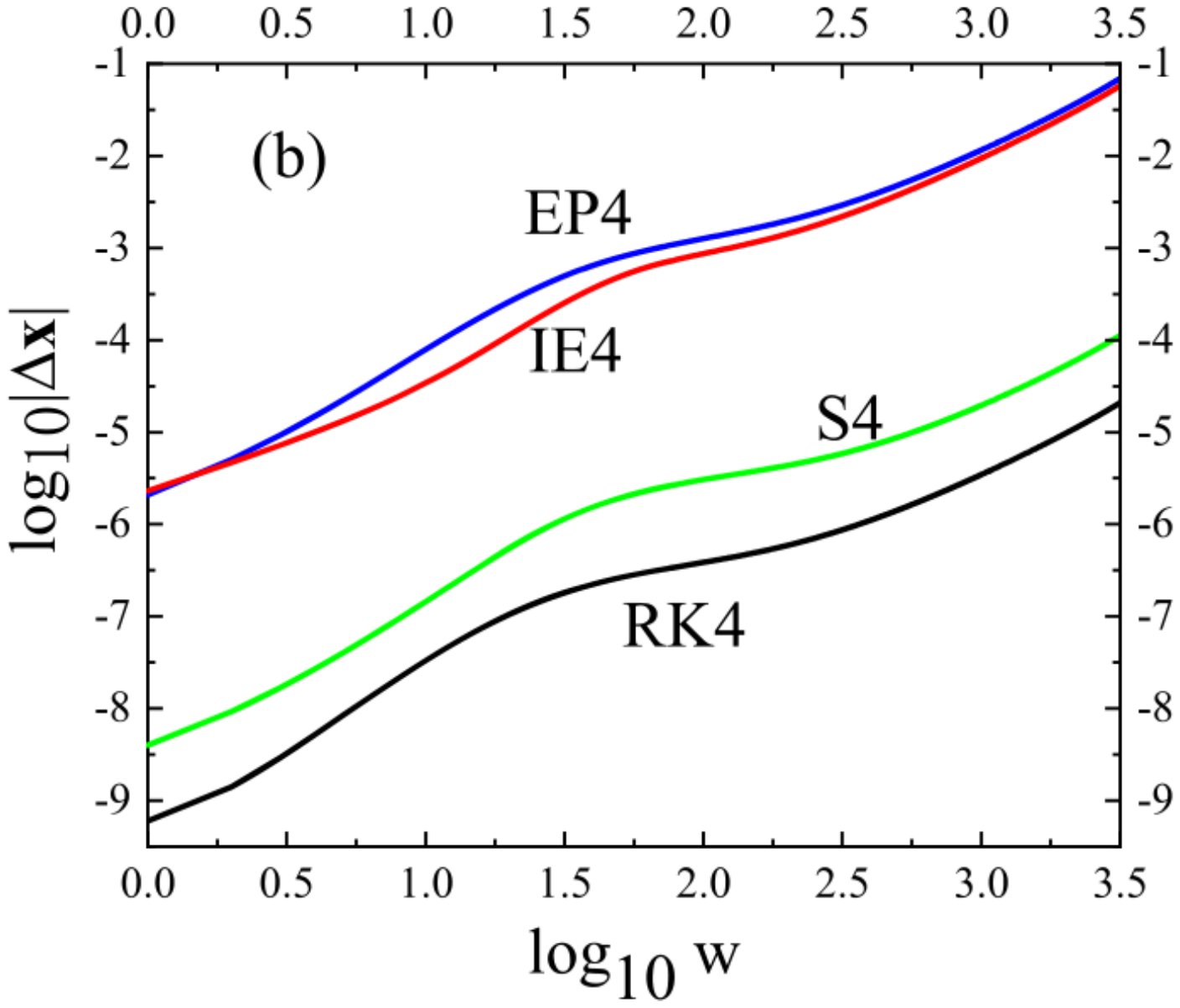}
\caption{ (a) Same as Figure 2, but the fourth-order methods show
the errors of the Carter constant for massive particles in
Equation (A3). The three methods S4, IE4 and EP4 conserve the
Carter constant, but RK4 does not. The notation RK4*0.01 means
that the plotted errors for RK4 are decreased 100 times, compared
with the realistic errors. IE4 is slightly better than S4 in
accuracy of the constant, but is slight poorer than EP4. (b)
Accuracy of the solution, estimated by $\Delta
\textbf{x}=\mathbf{\bar{x}}-\textbf{x}$, where
$\textbf{x}=(r,\theta, p_r, p_{\theta})$ is given by one of the
four algorithms, and $\mathbf{\bar{x}}$ is a high-precision
reference solution obtained from an eighth- and ninth-order
Runge-Kutta-Fehlberg integrator [RKF8(9)] with adaptive step
sizes. For an integration time of $w=2000$,  accuracy of the
solution is the best for RK4, whereas the poorest for EP4 or IE4.
}
 \label{Fig3}}
\end{figure*}

\begin{figure*}[ptb] \center{
\includegraphics[scale=0.18]{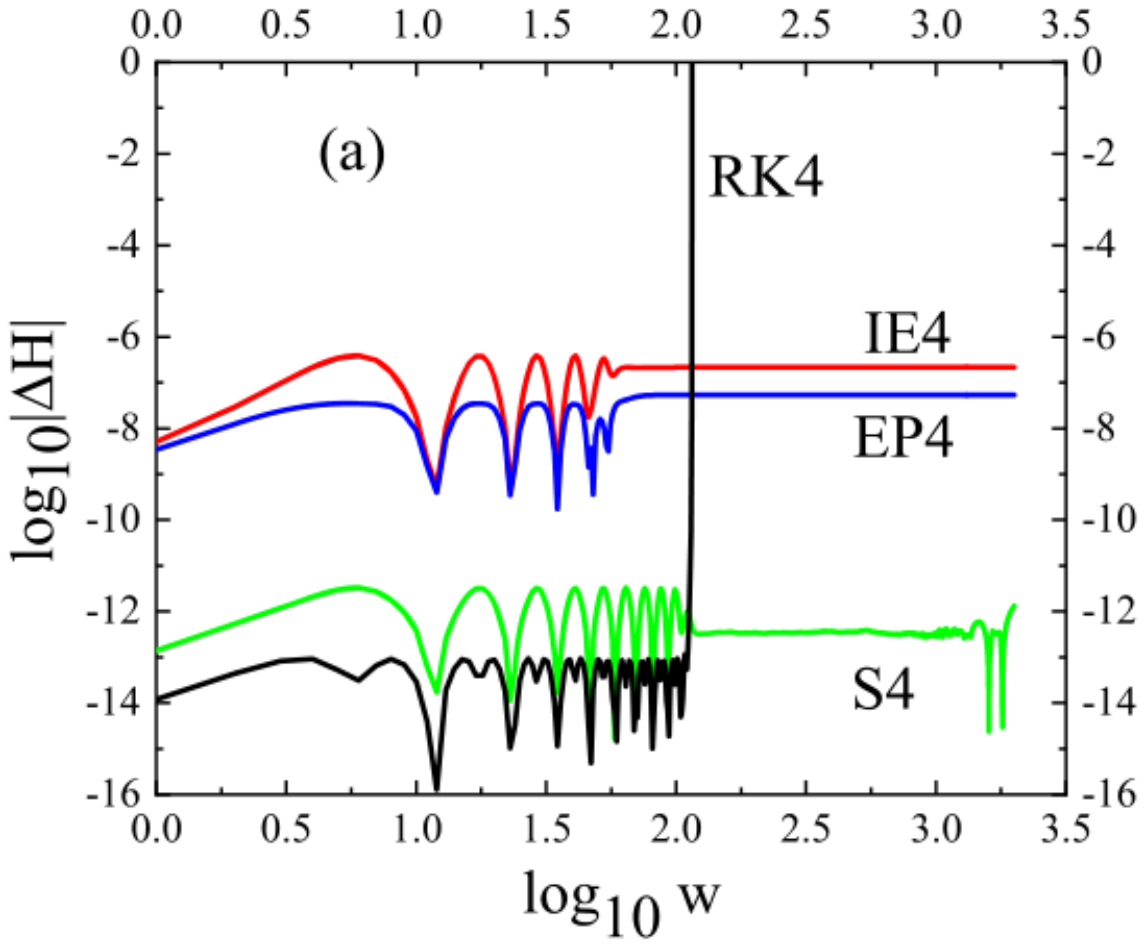}
\includegraphics[scale=0.18]{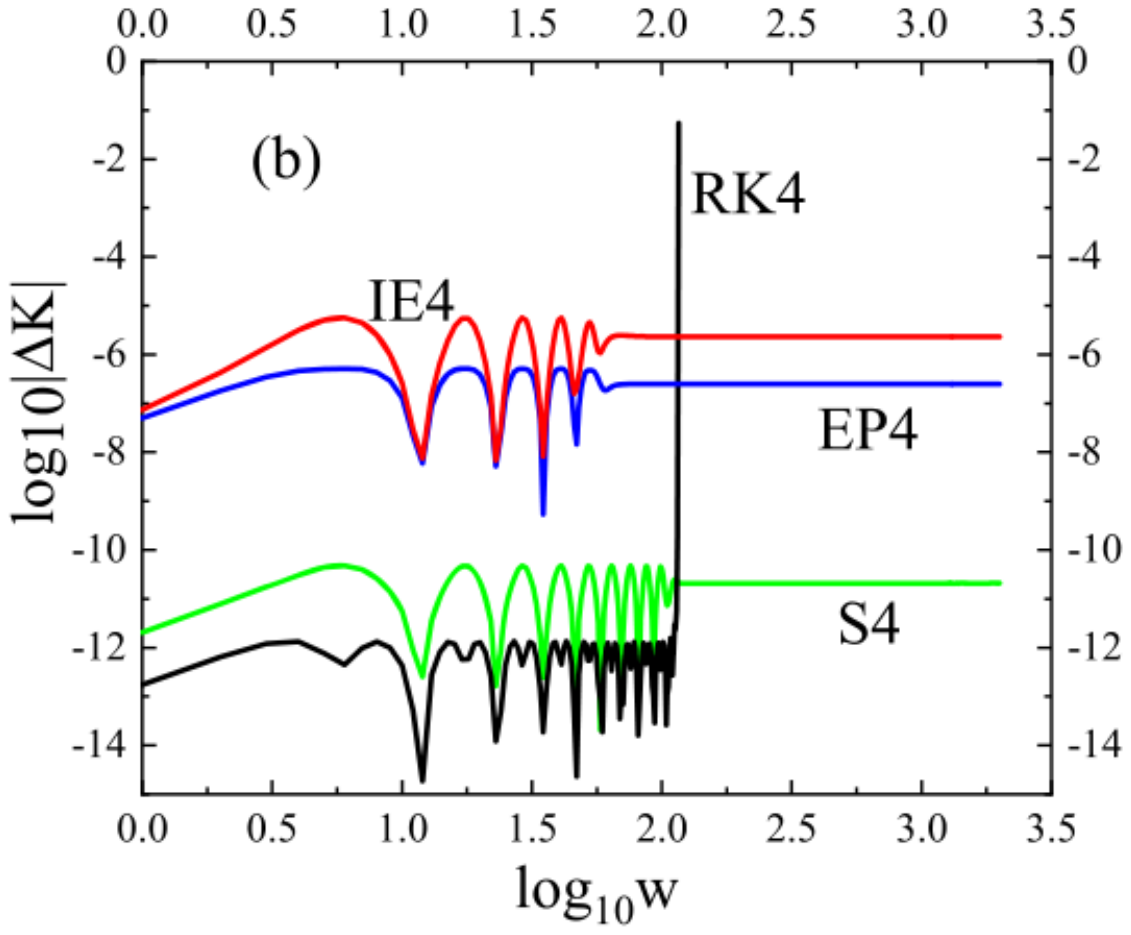}
\includegraphics[scale=0.18]{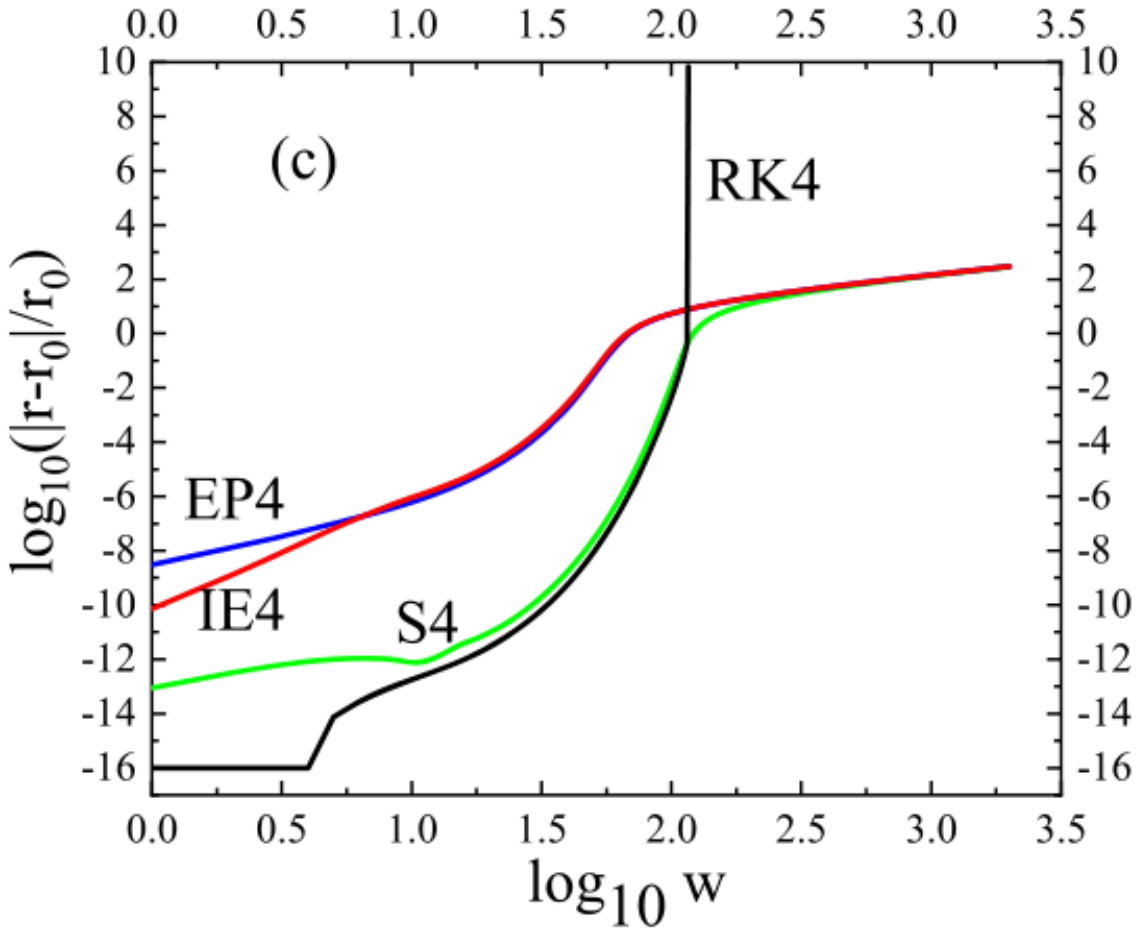}
\caption{ Tests of algorithmic performance for an unstable
spherical photon orbit. The new coordinate time step is $h=0.01$.
(a) Errors of the Hamiltonian. (b) Errors of the Carter constant.
(c) Relative errors of the radius. The errors of the Hamiltonian
and Carter constants are the worst when RK4 arrives at an
integration time of $w=116$. However, the two constants are
conserved by anyone of S4, IE4 and EP4 during an integration time
of $w=2000$. Unlike RK4, each of S4, IE4 and EP4 can work well in
conservation of the two constants although this orbit does not
remain spherical due to orbital instability after the integration
time $w=116$.}
 \label{Fig4}}
\end{figure*}

\begin{figure*}[ptb]
\center{
\includegraphics[scale=0.25]{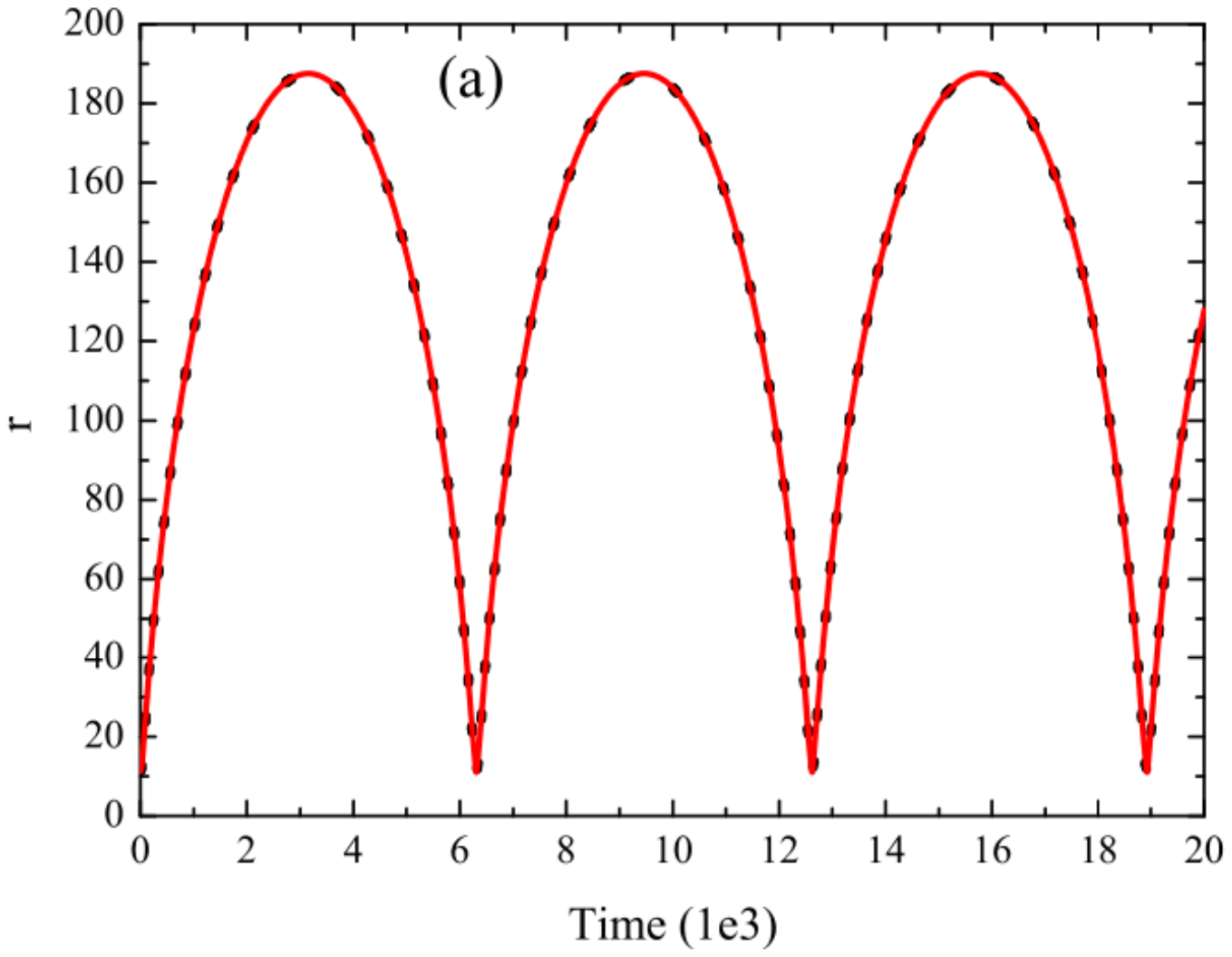}
\includegraphics[scale=0.25]{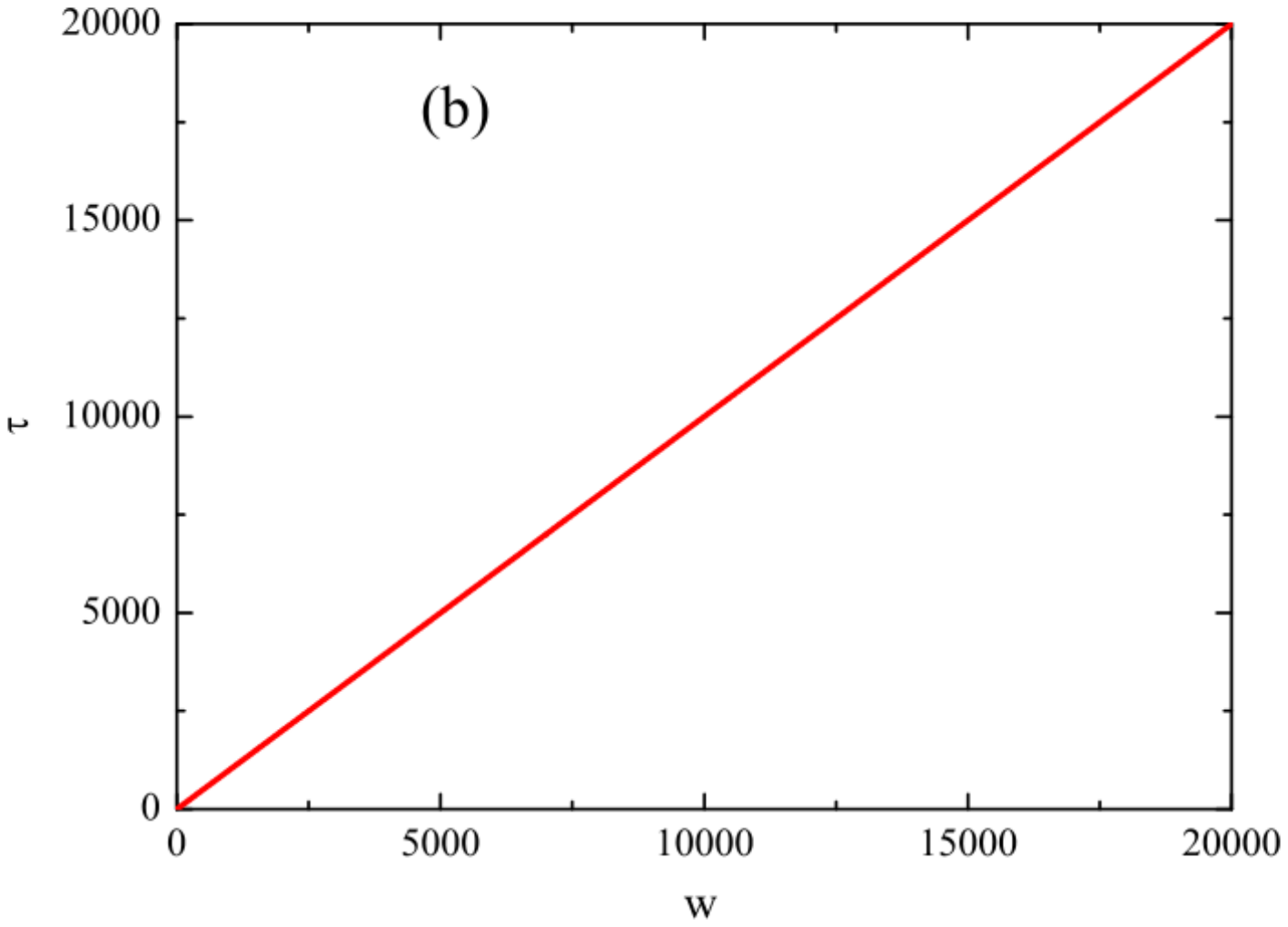}
\caption{(a) Evolution of $r$ with time for massive particles in
Figures 1 and 2. The red line corresponds to the evolution of $r$
with new coordinate time $w$, given by the fourth-order explicit
symplectic integrator S4 integrating the time-transformed
Hamiltonian (26) with new coordinate time step $h=1$. The black
dot relates to the evolution of $r$ with proper time $\tau$,
obtained from the fourth-order implicit and explicit mixed
symplectic method IE4 solving the original Hamiltonian system (18)
with proper time step $h=1$. The former evolution curve is almost
consistent with the latter one. (b) The relation between new
coordinate time $w$ and proper time $\tau$ for S4 integrating the
time-transformed Hamiltonian (26) with new coordinate time step
$h=1$. It is clear that the relation is $w=\tau$ if the
integration time is not long enough.}
 \label{Fig5}}
\end{figure*}

\end{document}